\documentclass[journal abbreviation, manuscript]{copernicus}

\usepackage{url} 

\usepackage{textcomp} 
\usepackage{subfig}

\usepackage{natbib} 
\bibpunct{(}{)}{;}{a}{,}{,~}

\usepackage{enumitem} 
\usepackage{float} 
\usepackage{url}
\usepackage{hyperref}
\hypersetup{colorlinks=true,citecolor=black,filecolor=black,linkcolor=black,urlcolor=black,breaklinks=true}

\newcommand\blfootnote[1]{%
	\begingroup
	\renewcommand\thefootnote{}\footnote{#1}%
	\addtocounter{footnote}{-1}%
	\endgroup
}

\usepackage{framed}
\usepackage{soul}

\begin{document}


\title{Carbon Cycle Extremes Accelerate Weakening of the 
Land Carbon Sink in the Late 21st Century}

\blfootnote{This manuscript has been authored by UT-Battelle, LLC, under contract DE-AC05-00OR22725 with the US Department of Energy (DOE). The US government retains and the publisher, by accepting the article for publication, acknowledges that the US government retains a nonexclusive, paid-up, irrevocable, worldwide license to publish or reproduce the published form of this manuscript, or allow others to do so, for US government purposes. DOE will provide public access to these results of federally sponsored research in accordance with the DOE Public Access Plan (http://energy.gov/downloads/doe-public-access-plan).}

\Author[1,2]{Bharat}{Sharma}
\Author[3]{Jitendra}{Kumar}
\Author[1]{Auroop~R.}{Ganguly}
\Author[2,4]{Forrest~M.}{Hoffman}

\affil[1]{Sustainability and Data Sciences Laboratory, Department of Civil and Environmental Engineering, Northeastern University, Boston, Massachusetts, USA}
\affil[2]{Computational Sciences \& Engineering Division and the Climate Change Science Institute, Oak Ridge National Laboratory, Oak Ridge, Tennessee, USA}
\affil[3]{Environmental Sciences Division, Oak Ridge National Laboratory, Oak Ridge, Tennessee, USA}
\affil[4]{Department of Civil and Environmental Engineering, University of Tennessee, Knoxville, Tennessee, USA}






\correspondence{Bharat Sharma (bharat.sharma.neu@gmail.com)}

\runningtitle{TEXT}

\runningauthor{TEXT}

\received{}
\pubdiscuss{} 
\revised{}
\accepted{}
\published{}


\firstpage{1}

\maketitle


\begin{abstract}

Increasing surface temperature could lead to enhanced evaporation, reduced soil moisture availability, and more frequent droughts and heat waves. 
The spatiotemporal co-occurrence of such effects further drives extreme anomalies in vegetation productivity and net land carbon storage. 
However, the impacts of climate change on extremes in net biospheric production (NBP) over longer time periods are unknown. 
Using the percentile threshold on the probability distribution curve of NBP anomalies, we computed negative and positive extremes in NBP. Here we show that due to climate warming, about 88\% of global regions will experience a larger magnitude of negative NBP extremes than positive NBP extremes toward the end of 2100, which accelerate the weakening of the land carbon sink. 
Our analysis indicates the frequency of negative extremes associated with declines in biospheric productivity was larger than positive extremes, especially in the tropics. 
While the overall impact of warming at high latitudes is expected to increase plant productivity and carbon uptake, high-temperature anomalies increasingly induce negative NBP extremes toward the end of the 21st century. 
Using regression analysis, we found soil moisture anomalies to be the most dominant individual driver of NBP extremes. 
The compound effect of hot, dry, and fire caused extremes at more than 50\% of the total grid cells.
The larger proportion of negative NBP extremes raises a concern about whether the Earth is capable of increasing vegetation production with growing human population and rising demand for plant material for food, fiber, fuel, and building materials. 
The increasing proportion of negative NBP extremes highlights the consequences of not only reduction in total carbon uptake capacity but also of conversion of land to a carbon source.

\end{abstract}

\section*{Short summary}

Rising atmospheric CO$_{2}$ enhances vegetation growth through increased carbon fertilization and water-use efficiency;
however, it also increases surface temperature, which could lead to enhanced evaporation, 
reduction in available soil moisture, and increases in droughts and heatwaves.
The impact of such climate extremes is detrimental to terrestrial carbon uptake capacity.
We investigated extreme events in net biospheric production (NBP), a component of the terrestrial carbon cycle and a measure of the uptake of carbon by plants.
We found that due to an overall climate warming, about 70\% of the world's regions towards the end of 2100 will show anomalous losses in biome productivity rather than gains.
While lack of soil moisture alone causes the largest number of losses in biome productivity, the compound effect of hot, dry, and fire events drives 50\% of all NBP extremes.
A few high latitude regions that were net carbon sinks during warm months are expect to show negative temperature sensitivity to NBP over time. 
This increased number of negative NBP extremes raises a concern about whether the Earth is capable of increasing its capacity 
to increase vegetation production with increasing human population and rising plant material per capita demand for food, fiber, fuel, and building material.



\section{Introduction}
\label{sec:intro}

Rising anthropogenic carbon dioxide (CO\textsubscript{2}) emissions are
leading to increases in Earth's surface temperature and climate variability
as well as intensification of climate extremes. Terrestrial ecosystems have
historically taken up a little over one-quarter of these
emissions via carbon accumulation in forest biomass and soils
\citep{friedlingstein_2019_GCB} and helped constrain increasing
atmospheric CO\textsubscript{2} concentrations. The increase in the net
terrestrial carbon sink is a result of reduced deforestation, enhanced vegetation
growth driven by CO\textsubscript{2} fertilization, and lengthening of
growing seasons in high latitudes. The growing terrestrial carbon sink provides
a negative feedback to climate change; however,
exacerbating environmental changes and climate extremes, such as droughts,
heatwaves and fires, have the potential to reduce regional carbon stocks 
and moderate carbon uptake \citep{Reichstein_2013_climate_ext, Sharma_2022_CarbonExtremesLULCC}.
Net biospheric production (NBP), the total downward flux of carbon from the atmosphere to the land,
represents the net carbon uptake after accounting for carbon losses 
from plant respiration, heterotrophic respiration, fire, and 
harvest \citep{Bonan_Book_2015} and is a critical measure of 
land carbon storage. Climate-driven
large anomalies in NBP could impact the structure, composition, 
and function of terrestrial ecosystems \citep{Frank_ClimateExtremes_CC_2015}. 
To improve our understanding of the climate--carbon cycle feedbacks, especially
during such large carbon anomalies, we investigated the changing
magnitude, frequency, and spatial distribution of NBP extremes
over decadal time periods and identify the influential climate anomalies
that potentially drive large NBP extremes at regional and global scales.

Terrestrial carbon cycle processes, such as photosynthesis, 
respiration and elemental cycling control the structure, composition and
function of terrestrial ecosystems.
In the past
few decades, the global terrestrial carbon cycle has taken up 25--35\%
of the CO\textsubscript{2} emissions from anthropogenic activities such
as fossil fuel consumption, deforestation and other land use changes \citep{Piao_2019_CC}.
With rising atmospheric CO$_2$, the
carbon uptake by both the land and the ocean has also increased but with
significantly greater variability over land \citep{friedlingstein_2019_GCB}. 
The interannual variability in land carbon uptake is strongly influenced by climate extremes, and it is primarily responsible for the interannual variation in the atmospheric CO$_2$ growth rate \citep{Piao_2019_CC}.

Climate extremes are part of Earth's climatic variability, affecting
terrestrial vegetation and modifying ecosystem-atmosphere feedbacks
\citep{buttlar_2018_climate_extremes}. 
Recent
studies have investigated the influence of rising temperatures on climate extremes and
terrestrial ecosystems \citep{buttlar_2018_climate_extremes,
Diffenbaugh_2017_climate_extremes,Frank_ClimateExtremes_CC_2015,
Zscheischler_Compound_2018, Sharma_2022_CarbonExtremesLULCC}. 
Observations
and climate models suggest that climate change has increased the
severity and occurrence of the hottest month, hottest day, and driest and
wettest periods \citep{Diffenbaugh_2017_climate_extremes}. 
Heavy precipitation or lack thereof could have a negative
feedback on the carbon cycle via soil water-logging or drought stress,
respectively \citep{Reichstein_2013_climate_ext}. 
A few studies
have investigated the impact of climate extremes on the carbon cycle and
found that hot and dry extremes reduce land carbon uptake, especially in
low latitudes and arid/semi-arid regions
\citep{Pan_2020_CarbonClimateExtremes,Frank_ClimateExtremes_CC_2015}.
Attribution studies infer that the compound effect of multiple
climate drivers has a larger effect on the carbon cycle and its extremes
\citep{Sharma_2022_CarbonExtremesLULCC, Zscheischler_Compound_2018, Pan_2020_CarbonClimateExtremes,
Frank_ClimateExtremes_CC_2015, Reichstein_2013_climate_ext} than any
individual climate driver.  
Most
attribution methods focus on analyzing the response of the carbon cycle to
climate, aggregated over annual, sub-annual, and seasonal scales;
however, the responses may vary over shorter time scales, including daily to monthly.

The variability in climate--carbon cycle feedbacks is dependent on geographical
location, among other factors. \citet{Grose_2020_pr_australia} reported
that while Australia is expected to experience an overall reduction in
precipitation by 2100, the spatial distribution of precipitation varies since a few 
regions are expected to get more
and others will receive less precipitation. \citet{Ault_2020_droughts}
found that despite the overall increase in precipitation and water-use
efficiency globally, the available soil moisture may be reduced across many
regions due to increased evapotranspiration from higher temperatures,
exceeding the supply from precipitation. The regions that see a decrease
in supply and increase in demand for water are sensitive even to
relatively small increases in temperature. These feedbacks will increase the
severity of droughts, and ENSOs may further amplify the effect
\citep{Ault_2020_droughts}. Net primary production (NPP)
sensitivity to temperature is negative above 15$^{\circ}$C and
positive below 10$^{\circ}$C \citep{Pan_2020_CarbonClimateExtremes},
which means warming will cause a reduction in carbon uptake in the tropics
and extra-tropics and an increase in carbon uptake at higher latitudes.
However, with increasing average surface temperatures, the NPP
sensitivity could become negative over time in high latitude regions.

Rising atmospheric CO\textsubscript{2} levels and climate change could have
implications for biological \citep{Frank_ClimateExtremes_CC_2015} and
ecological systems since the severity and occurrence of climate extremes,
such as heatwaves, droughts, and fires, are likely to strengthen in the 
future. These systems are more sensitive to climate and carbon
extremes than to gradual changes in climate.
The increasing frequency and magnitude of climate
extremes could reduce carbon uptake in
tropical vegetation, reduce crop yields \citep{Ribeiro_2020_Crop}, and
negate the expected increase in carbon uptake
\citep{Reichstein_2013_climate_ext}. 
In this study we investigated the extremes in NBP and their climate
drivers from Earth system model simulations for the period 1850--2100 
across several regions around the globe.
The objectives of this research were to
1) quantify the magnitude, frequency, and spatial distribution of NBP extremes, 
2) attribute individual and compound climate drivers of NBP extremes at multiple time lags, 
and 3) investigate the changes in climate--carbon cycle feedbacks at regional scales.

\section{Methods}
\label{sec:methods_begin}

\subsection{Data}

We used the Community Earth System Model (version 2) (CESM2) simulations
at 1$^{\circ}$\,$\times$\,1${^\circ}$ spatial and monthly temporal resolution 
to analyze the carbon cycle extreme events in net biospheric production.
The CESM2 is a fully
coupled global Earth system model composed of atmosphere, ocean, land, sea ice and land ice
components. 
The simulations analyzed here were forced with atmospheric greenhouse gas 
concentrations, aerosols and land use change for the historical
(1850--2014) and Shared Socioeconomic Pathway~8.5 (SSP5-8.5; 2015--2100)
scenario, wherein atmospheric CO\textsubscript{2} mole fraction rises from about 280~ppm in
1850 to 1150~ppm in 2100 \citep{Danabasoglu_2020_CESM2}.
While
CO\textsubscript{2} forcing causes temperatures to increase, changes in aerosols and land use
have a slight cooling effect \citep{Lawrence_2019_CLM5},
resulting in an overall increase of about $8^{\circ}$C in mean air
temperature over the global land surface during 1850--2100. 
All the variables used in this study are from the CESM2 simulation outputs.

\subsection{Definition and Calculation of Extreme Events}
\label{sec:methods_extremes}

We wanted to quantify the NBP extremes that are significant globally and compare  
the distribution of global NBP extremes across various regions. 
The Intergovernmental Panel on Climate Change (IPCC) \citep{Seneviratne_IPCC_2012} 
defines extremes of a variable as the subset of values in the tails of the 
probability distribution function (PDF) of anomalies.
Based on the global PDF of NBP anomalies, we selected a threshold value of ${q}$, 
such that total positive and negative extremes constitute 5\% of all NBP 
anomalies (Figure~\ref{f:schematic_extremes}).
The negative and positive extremes in NBP were comprised of NBP anomalies smaller 
than ${-q}$ and larger than ${q}$, respectively. 
While the total number of NBP extremes was constant (i.e., 5\% of all NBP anomalies) 
for any time period, the count and intensity among positive and negative extremes vary 
depending on the nature of the PDF of NBP anomalies. 

We computed extremes for every 25-year period from 1850 through 2100 to analyze the changing 
characteristics of NBP extremes at regional to global scales.
For regional analysis, we used the 26 regions defined in the IPCC Special Report on Managing
the Risks of Extreme Events and Disasters to Advance Climate Change Adaptation 
\citep{Seneviratne_IPCC_2012}, hereafter referred to as the SREX regions (Figure~\ref{f:srex_regions}).
We analyzed the characteristics of NBP extremes during carbon uptake periods,
when photosynthesis dominates NBP, and land is a net sink of carbon (NBP $> 0$), 
and carbon release periods, when NBP is dominated by respiration and disturbance 
processes, and land is a net source of carbon (NBP $< 0$) \citep{Marcolla_2020_NBP}.

The anomalies in NBP were calculated by removing the modulated annual cycle and nonlinear trend from the time series of NBP at every grid cell. 
We calculated the modulated annual cycle and nonlinear trend of NBP using singular spectrum analysis, which is a non-parametric spectral estimation method that decomposes a time series into independent and interpretable components of predefined periodicities \citep{Golyandina_Book_20010123}.
The conventional way of computing annual cycle or climatology does not capture the intrinsic non-linearity of the climate--carbon feedback \citep{Wu_MAC_2008}. 
The modulated annual cycle, composed of signals with return periods of 12 months and its harmonics, is able to capture the varying modulation of the seasonality of NBP under rising atmospheric CO\textsubscript{2} concentrations.
The nonlinear trend is comprised of return periods of 10 years and longer, such that the anomalies in the ecosystem and climate drivers capture the effects of ENSO, which has a large impact on climate and the carbon cycle \citep{Zscheischler_2014_GPP_extremes, Ault_2020_droughts}. 
Thus, NBP anomalies consist of intra-annual variability, represented by high-frequency signals (\textless12~months), and the interannual variability (\textgreater12~months and \textless10~years).

\subsection{Attribution to Climate Drivers}
\label{sec:methods_attribution}

While most studies traditionally attribute carbon cycle impacts to changes in climate
at seasonal to annual time scales, many carbon cycle responses to climate variability 
occur at shorter, daily to monthly, time scales \citep{Frank_ClimateExtremes_CC_2015}. 
Attribution analysis was performed in recent studies of large connected manifolds of 
spatio-temporal continuous extremes in the carbon cycle \citep{Sharma_2022_ICDM, Zscheischler_2014_GPP_extremes, Flach_Climate_extreme_GPP_2020} by 
comparing the medians or mean state of climate driver(s)
during or preceding a carbon cycle extreme with climate extreme for the same 
spatio-temporal region or grid cells affected during carbon cycle extremes. 
This method may not capture the variability at smaller regional to grid cell scales.
As described by \citet{Sharma_2022_CarbonExtremesLULCC}, we compute time-continuous extreme (TCE) 
events at every grid cell provided they fulfill the following conditions, 
(i) they must consist of isolated extremes that are continuous for at least one season (i.e., 3 months) 
and (ii) any number of isolated or contiguous extremes can be a part of a TCE event 
if the gap between such extremes is less than one season in length (i.e., up to 2 months).
We assume that extreme events separated by gaps greater than or equal to one season length are separate TCE events.
Using linear regression of time-continuous NBP extremes that represent the 
large intra-annual and interannual variation in NBP with climate anomalies, we 
quantified the dominance (regression coefficient) and response (sign of 
the regression coefficient) of climate drivers on large NBP extremes.

Human activities, such as fossil fuel emissions and land use changes, modify biogeochemical and biogeophysical processes, which alter the climate through climate--carbon feedbacks. 
Large anomalous changes in climate drivers have a strong impact on carbon uptake and biospheric productivity. 
Here, we attributed NBP TCEs to climate drivers, namely, precipitation (``Prcp''), soil moisture (``SM''), 
monthly average daily temperature (``TAS''), and biomass loss from carbon flux into the atmosphere due to 
fire (``Fire''). 
As the terrestrial vegetation has ingrained plasticity to buffer and push back effects of climate change \citep{Zhang_Lag_effects_2014}, the impacts of changes in climate drivers are often associated with lagged responses.
Moreover, the strength of the impact of climate on NBP is dependent on location, timing, and land cover type \citep{Frank_ClimateExtremes_CC_2015}. 
Linear regressions of TCEs in NBP and anomalies of every climate driver were identified at all affected land grid cells for lags from one to four months. 
We assumed that higher the Pearson Correlation coefficient (\(\rho\)) of a climate driver with NBP extremes, larger is its impact on NBP at that location. 
The attribution based on \(\rho\) is used only for those grid cells for which the significance value $p$~\textless~0.05. 
The grid cells with at least two negative and positive NBP TCEs each
often yielded high correlation coefficients with high significance values
($p<0.05$); thus, this constraint was applied for attribution to climate drivers.

The instantaneous impact (when lag equals zero months) of driver anomalies ($dri_t$)
on NBP TCEs ($nbp_t$) is computed using Equation~\ref{eqn:lag0}.
Attribution based on the lagged response of driver anomalies 
on NBP TCEs was computed using Equation~\ref{eqn:lagGT0}, where $L$ represents 
the number of lagged months. 
For lags greater than one month, we computed the correlation of the
average of climate driver anomalies, $\frac{dri_{t - l}}{L}$ 
for every time-step in the driver anomalies, with \textit{nbp$_{t}$}.
The resulting \(\rho\) captures the average response of antecedent climatic 
conditions up to \textit{L} months in advance that drive NBP TCEs. 

\hfill

\indent   for lag = 0:\\
\begin{equation}\label{eqn:lag0}
 \rho = corr(dri_{t}, nbp_{t}) 
\end{equation}


\indent   for lag \textgreater 0: \\
\begin{equation}\label{eqn:lagGT0}
 \rho = corr(\sum_{l=1}^{l=L}\left (\frac{dri_{t - l} }{L} \right ), nbp_{t})  
\end{equation}

\hfill

The direction and strength of the impact of various climate drivers on
plant productivity and the carbon sink vary with space and time. 
Increased temperatures could lead to increased respiration and 
losses in NBP in the tropics and mid-latitudes, but an increase in 
temperature could conversely lead to higher photosynthetic activity and greater uptake at higher 
latitudes. 
A moderate reduction in precipitation may not severely impact
vegetation productivity, but if accompanied by a heatwave, it could lead 
to large losses in NBP. 
We expect that CESM2 could simulate the impact of variability of climate drivers on 
ecosystem processes because the Community Land Model 
version 5 (CLM5 \citep{Lawrence_2018_CLM5_TechDoc}),
the land model component 
of CESM2, simulates water exchange across the root structure that varies with 
soil depth and plant functional type.
The soil water flux is dependent on hydraulic conductivity and hydraulic 
potential among various soil layers via Darcy's Law.
Due to the differences in hydraulic properties of soil layers, 
their soil water content varies by soil depth. 
The root-soil conductivity depends on evaporative demand and varies by 
soil layer and is calculated based on soil potential and soil properties, 
via Brooks-Corey theory. 
The rooting depth parameterizations were improved in CLM5 with a deepened rooting profile for broadleaf evergreen and broadleaf deciduous tropical trees \citep{Lawrence_2019_CLM5}.

Anomalous climate drivers causing NBP extremes may or may not 
qualify as climate extremes by themselves. 
A recent study found that the periods of extreme climate and NBP 
often do not occur at the same time, and the compound effect of non-extreme 
climate drivers could produce an extreme in NPP \citep{Pan_2020_CarbonClimateExtremes}. 
Occurrence of a NBP extreme is also likely driven by the compound effect of
multiple climate divers, we identified co-occurring anomalous climatic 
conditions during and antecedent to NBP extremes to improve our 
understanding of the interactive compound effect of drivers on the carbon cycle.
We analyzed the dominant climate drivers across SREX regions for every 
25-year period from 1850 to 2100 to understand the changing 
characteristics of large spatio-temporal extremes and their compound climate 
drivers across time and space.
Since the dominance of climate drivers are usually quantified by a correlation 
coefficient range of 0.5--0.7 \citep{Dormann_2013_Collinearity}, we imposed a limit of 
correlation coefficients $\rho > 0.6$ and significance 
values $p < 0.05$ on co-occurring individual climate drivers to 
qualify as individual or compound drivers of NBP extremes.
These constraints yield a smaller number of extremes that are attributable to 
climate drivers with high confidence.

\section{Results and Discussion}
\subsection{Characteristics of NBP Extremes}
\label{sec:results_global_extremes}

\begin{figure}
\centering
 \includegraphics[trim = {0.1cm 0.1cm 0.1cm 0.1cm}, clip,width=0.95\columnwidth]{
    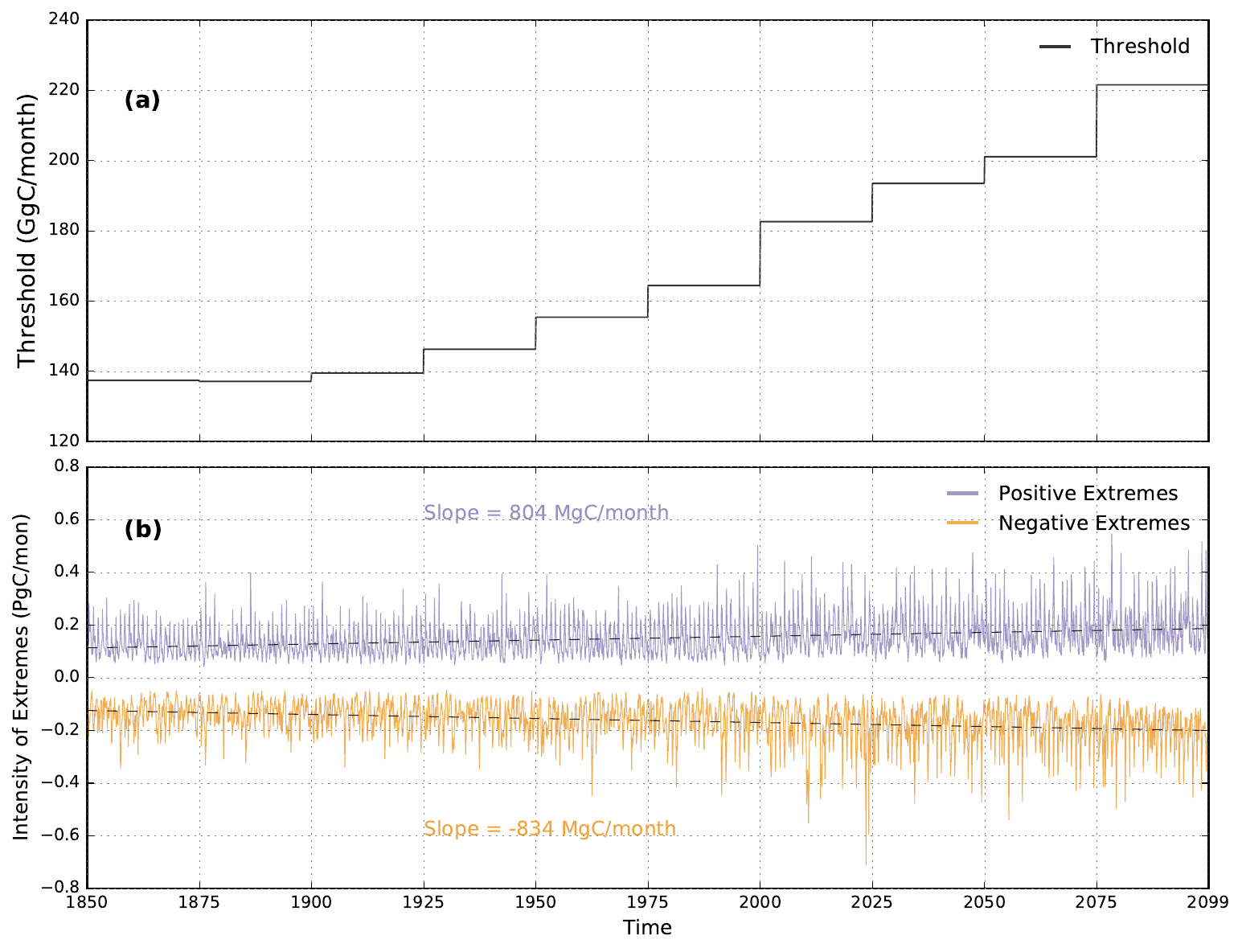}
\caption{ (a) The 5\textsuperscript{th} percentile threshold, \textit{q}, of NBP anomalies. 
    The negative extremes in NBP are those NBP anomalies 
    that are $<-q$ and positive extremes are $>q$.
    (b) The intensity of positive and negative extremes in NBP in CESM2 are represented by green and 
    red color, respectively. 
    The rate of increase of positive and negative extremes in NBP are 
    804 and $-$834~MgC per month, respectively.}
\label{f:nbp_cesm2_intensity}
\end{figure}

The 5\textsuperscript{th} percentile NBP anomalies computed for every 25-year period from 1850 to 2100 rendered 
threshold trajectories that increase from 140~GgC/month to 220~GgC/month (Figure~\ref{f:nbp_cesm2_intensity}(a)). 
This 1.5 times increase in threshold values demonstrates the increasing magnitude of anomalies and interannual variability of NBP across the globe. 
The corresponding time series of intensity of losses and gains in biome
productivity were calculated by integrating the negative (NBP anomalies
		$<-q$) and positive (NBP anomalies $>q$) extreme anomalies.
The rate of increase in the magnitude of negative extremes 
($-$834~MgC/month) was larger than that of the positive extremes (804~MgC/month) (Figure~\ref{f:nbp_cesm2_intensity}(b)), which implies that over time the net losses in carbon storage during NBP extremes increases.

\begin{sidewaysfigure}
 \centering
 \includegraphics[trim = {.1cm .1cm 0.1cm 1.2cm}, clip,width=0.95\columnwidth]{
    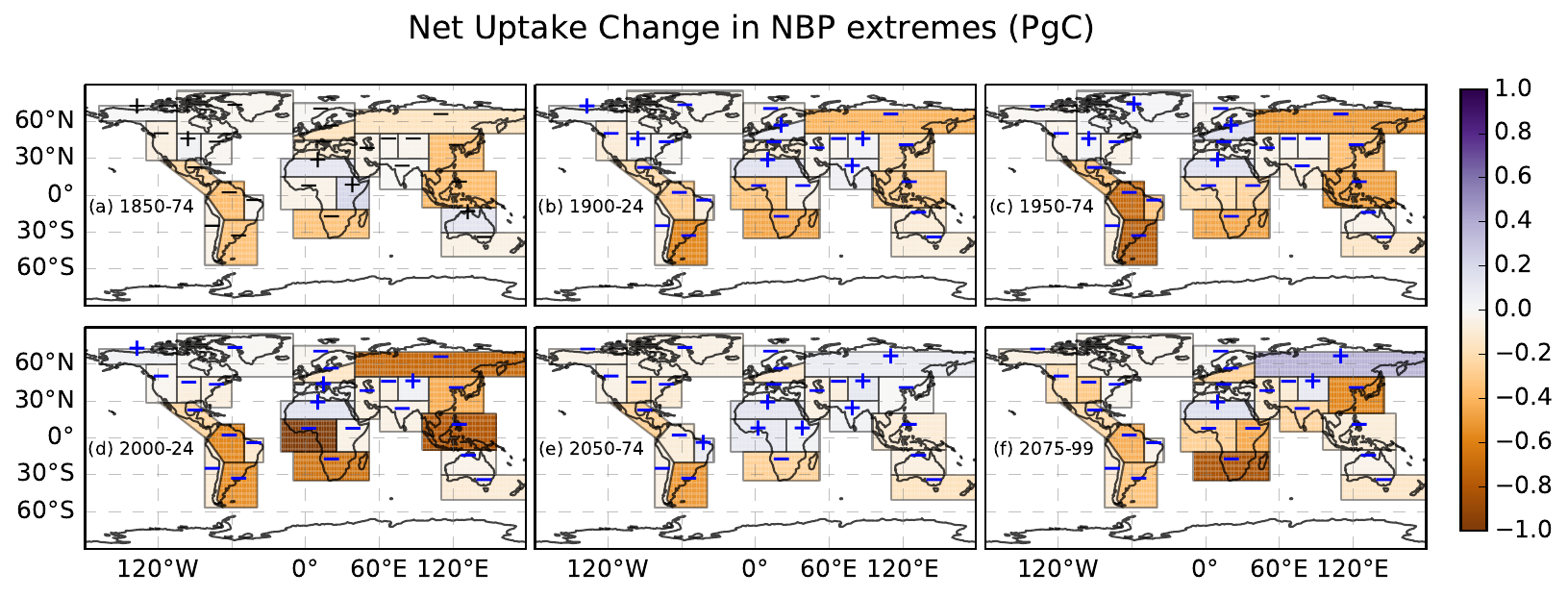}
 \caption{The figure shows the sum of the magnitude of positive and negative NBP extremes during 25 year periods. 
 The figure shows the total integrated net impact of carbon cycle extremes (PgC) across SREX regions for the following periods: (a) 1850--74, (b) 1900--24, (c) 1950--74, (d) 2000--24, (e) 2050--74, and (f) 2075--99. 
 A net gain in carbon uptake during extremes is represented by a purple color and a `+' sign, and a net decrease is represented by an orange color and a `$-$' sign. 
 For most regions, the magnitude of negative NBP extremes or losses in carbon uptake were higher than positive NBP extremes or gains in carbon uptake.
 }
\label{f:cesm2_nbp_net_uptake_reg}
\end{sidewaysfigure}

The changes in NBP are driven by spatial and temporal variations in climate drivers and anthropogenic forcing. 
During 1850--1874, 24 out of 26 regions were dominated by carbon release (Figure~S4) and 
the total NBP was negative (Figure~\ref{f:nbp_cesm2_rm5yr}).
From 1850 through the 1960s, the land experienced a net C losses in carbon storage flux
likely driven by deforestation, fire and land-use change 
activities \citep{friedlingstein_2019_GCB} (Figure~\ref{f:ts_carbonfluxes_tropics}).
After the year 1960, the continued increase in fossil fuel emissions raised the atmospheric CO\textsubscript{2} concentration despite declining rates of deforestation.
Increasing CO\textsubscript{2} fertilization, water-use efficiency, and
the lengthening of growing seasons enhanced vegetation growth and 
NBP with the largest increases in the tropics and northern high latitudes (Figure~\ref{f:cesm2_nbp_reg_total}). 
After 2070, the total NBP reached its peaks and started to decline (Figure~\ref{f:nbp_cesm2_rm5yr}) 
as ecosystem respiration exceeded total photosynthetic activity.
Tropical regions have the largest magnitude of NBP; however, the 
rate of increase of NBP declined after 2050 and the region of 
Sahara (SAH) showed an early drop in total NBP after the year 2050.
Longer dry spells and intense rains due to changing precipitation patterns 
in Mediterranean and subtropical ecosystems are likely to cause 
higher tree mortality \citep{Frank_ClimateExtremes_CC_2015}. 
Hot temperatures and reduced activity of RuBisCO hindering 
carboxylation are possible factors that will likely cause a net 
decrease in NBP in the region of SAH and make it a net carbon source after 2050. 
During 2050--74 and 2075--99, low-latitude regions exhibit the highest regional NBP; 
however, many areas in the tropics exhibited a declining 
growth rate of regional NBP (Figure~\ref{f:cesm2_nbp_reg_total}).

\begin{figure}
 \centering
 \includegraphics[trim = {0.1cm 0.1cm 0.1cm .15cm},clip,width=0.95\columnwidth]{
    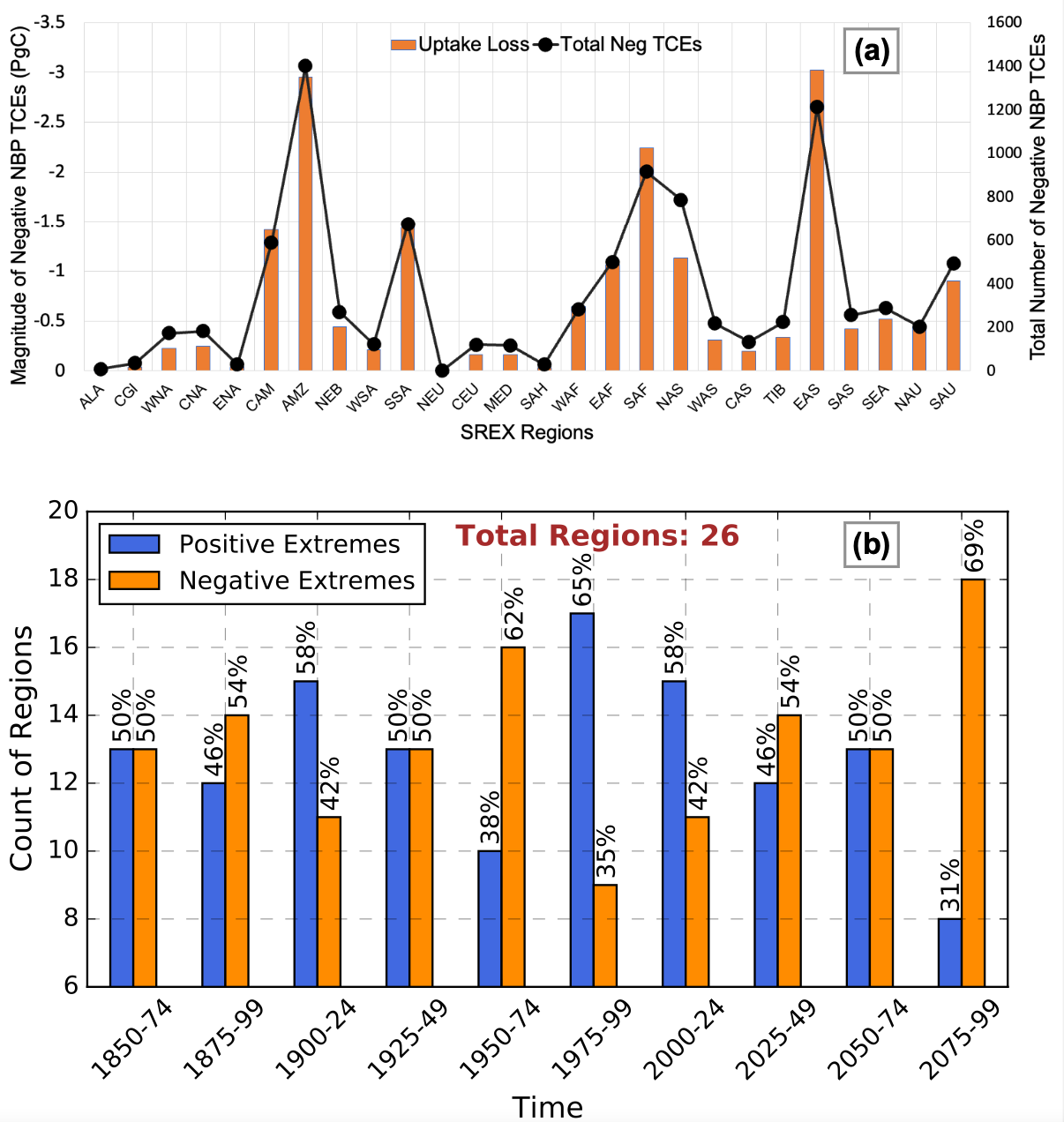}
 \caption{(a) Total magnitude of negative carbon cycle extremes or loss is carbon uptake during TCEs across SREX regions plotted as a bar graph (left $y$-axis). 
 The total number of negative TCE events (right $y$-axis) plotted as line graph. 
 The largest portion of carbon uptake loss is in the tropical SREX regions of the Amazon (AMZ), East Asia (EAS), and South Africa (SAF) for the period 2075--99. 
 (b) Count ($y$-axis) of the regions dominated by either positive or negative NBP extremes. 
 Relative to a total of 26 SREX-regions, the percent count of positive or negative NBP extremes is represented at the top of the bars.}
\label{f:Regions_NBP_TCE_extremes}
\end{figure}

As anomalous changes in climate vary over space and time, extremes in NBP also respond to the interactive effects of climate drivers and exhibit spatial and temporal variation. 
Figure~\ref{f:cesm2_nbp_net_uptake_reg} shows the net total sum of both positive and negative extremes in NBP in SREX regions integrated for all 25-year periods (1850--74, 1900--24, 1950--74, 2000--24, 2050--74, and 2075--99).
Most regions exhibited net losses in biospheric productivity during extremes, e.g., South Africa (SAF) has always been dominated by negative NBP extremes. 
The large magnitude of net carbon uptake changes during the period 2000--24 was likely due to the change in LULCC forcing from decadal to annual during 2000--2015 and then back to decadal from 2015 onward.
The increased temporal resolution of LULCC forcing possibly caused higher climate variability due to biogeophysical feedbacks and subsequently led to increased carbon cycle variability and extremes. 
Since we focused on NBP extremes, which are tails of PDF of anomalies (or interannual variability) of NBP, the magnitude of carbon cycle extremes was large during this period. 
However, the impact of LULCC forcing was not as significant on mean NBP changes (Figure S3).
23 out of 26 SREX regions experienced an overall loss in biospheric productivity during extremes near the end of the 21\textsuperscript{st} century (Figure~\ref{f:cesm2_nbp_net_uptake_reg}). 
The distribution of the total magnitude and count of negative TCEs during 2075--99 across all the 
SREX regions followed a similar pattern, i.e., more frequent extremes were accompanied by larger carbon 
losses (Figure~\ref{f:Regions_NBP_TCE_extremes}(a)). 
The largest losses in carbon uptake during TCEs were in tropical regions, 
e.g., East Asia (EAS), Amazon (AMZ), and SAF, with $-$3, $-$3, and $-$2.25~PgC carbon losses, respectively, during 2075--99. 
These regions also witnessed the highest number 
of negative NBP TCEs at 1270, 1410, 950, respectively.
The magnitude of carbon losses and the number of negative NBP TCEs were highest in tropical regions.
The magnitude and number of negative TCEs were very low for the high
latitude regions of Alaska (ALA), Canada and Greenland (CGI), Eastern North America (ENA), Northern Europe (NEU), Central Europe (CEU), 
and the dry regions of the Mediterranean (MED) and Sahara (SAH). 
Although the number of NBP TCEs in NAS were more than in Southeastern South America (SSA) 
and Central America (CAM), the magnitude of NBP TCEs in NAS were low due to lower regional NBP.
Since the extremes were calculated based on global anomalies, the largest impacts on 
terrestrial carbon uptake are expected in the regions of AMZ, EAS, and SAF, which have the largest concentrations of live biomass.



The magnitude and the total number of regions dominated by negative extremes in 
NBP are expected to gradually increase in the 21\textsuperscript{st} century 
(Figure~\ref{f:Regions_NBP_TCE_extremes}(b)). 
Most of the increase in the frequency of negative extremes in NBP are expected 
in ENA, South Asia (SAS), SAF, and CAM (Figure~\ref{f:Spatial_NBP_fq_pos_neg_ext}). 
The increase in the magnitude (23 out of 26 or 88\% of all regions) and frequency 
(18 out of 26 or 70\% of all regions) of negative NBP TCEs in most SREX regions 
during 2075--99 is a matter of concern since the total global NBP peaked at 
around 2070 and subsequently declines in the model (Figure~\ref{f:nbp_cesm2_rm5yr}). 
The negative NBP TCEs dominate in eight out of the nine tropical regions, 
which store the largest standing carbon biomass and represent the 
largest portion of carbon uptake loss during negative NBP extremes.
A large magnitude of extreme events in the NBP could potentially lead to a 
state of low and decreasing carbon sink capacity that could further lead to a
positive feedback on climate warming.
The strengthening of negative extremes relative to positive extremes in NBP represents
a net decline of terrestrial carbon sink capacity into the future (Figure~\ref{f:nbp_cesm2_rm5yr}).
This is contrary to the results of \cite{Zscheischler_2014_GPP_extremes}, who
found a strengthening of positive (net ecosystem production, NEP) extremes 
over time using CMIP5 ESMs. 
However, the ratio of negative to positive carbon cycle extremes 
in our study lie within the multi-model spread of the relative strength of NEP extremes 
in CMIP5 ESMs \citep{Zscheischler_2014_GPP_extremes}. 
The positive feedback of warming and climate driven losses in carbon uptake 
raise concerns about the implications of reducing terrestrial 
uptake capacity on food security, global warming, and ecosystem functioning.
Moreover, the sensitivity of vegetation responses is higher for climate and 
carbon extremes than for gradual changes because of larger response 
strength and shorter response times \citep{Frank_ClimateExtremes_CC_2015}. 

\subsection{Attribution to Climate Drivers} 
\label{sec:results_attribution}

\begin{sidewaysfigure}
 \centering
 \includegraphics[trim = {0.1cm 0.1cm 0.1cm .85cm},clip,width=0.9\columnwidth]{
    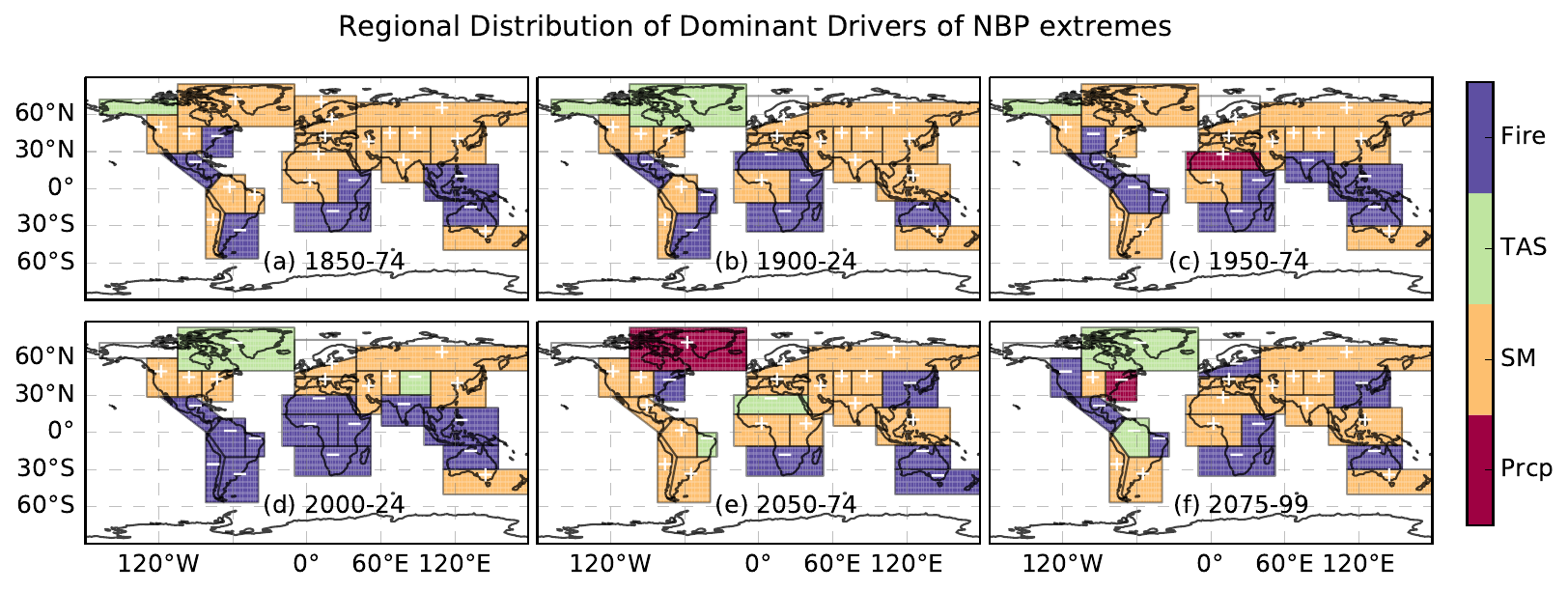}
 \caption{Spatial distribution of dominant climate drivers across SREX regions. The color in every SREX region represents the most dominant climate driver causing carbon cycle extremes at 1 month lag for following periods: (a) 1850--74, (b) 1900--24, (c) 1950--74, (d) 2000--24, (e) 2050--74, and (f) 2075--99. The positive (`$+$') and negative (`$-$') sign within a region represents the correlation relationship of NBP extremes with every dominant climate drivers.}
\label{f:cesm2_nbp_percent_dom_drivers_regional}
\end{sidewaysfigure}

The increase in climate variability and extremes driven by rising CO\textsubscript{2} 
emissions influences the terrestrial carbon cycle \citep{buttlar_2018_climate_extremes, 
Reichstein_2013_climate_ext, Sharma_2022_CarbonExtremesLULCC}. 
The control of climate drivers on NBP extremes is dependent on the regional interannual 
variability of climatic conditions and vegetation composition. 
The percent of total number of grid cells that show soil moisture as a dominant 
driver of NBP TCEs were about 40 to 50\% from 1850 to 2100 across 
multiple lags, which means that the near term and long term impact of soil 
moisture were highest among all other drivers (Figure~\ref{f:cesm2_nbp_percent_dom_drivers}). 
The positive response of soil moisture anomalies on NBP TCEs indicates that 
a decline in soil moisture causes a reduction in NBP and vice-versa. 
Likewise, the dominant climate driver across the 26 SREX regions was also 
soil moisture and it exhibited a positive response relationship with NBP TCEs 
(Figure~\ref{f:cesm2_nbp_percent_dom_drivers_regional}).  
However, the proportion of the total number of grid cells dominated by precipitation 
doubled (10 to 20\%) when the lag was increased from 1 to 3 months.
This implies that antecedent declines in precipitation limit carbon uptake 
more than a recent decline in precipitation and possibly causes a decline in soil moisture.
Moreover, the plants with deep roots are less impacted with 
short-term reduction in precipitation than prolonged droughts, which are 
caused by soil moisture limitation. 
By the end of the 21\textsuperscript{st} century, the model indicates that 
70\% of the total 
number of NBP extremes will be water-driven, i.e., due to 
soil moisture and precipitation.
Our results are consistent with recent studies 
\citep{liu_2020_SM_dryness,Pan_2020_CarbonClimateExtremes}
that concluded 
that the most important factor limiting vegetation growth is water 
stress, which is caused mainly by low soil moisture.
Lack of soil moisture for extended periods could result in drought events, 
causing a larger reduction in ecosystem productivity and a smaller reduction 
in terrestrial respiration. 

The second most dominant driver of NBP TCEs was fire, which has a positive 
response on NBP TCEs (Figure~\ref{f:cesm2_nbp_percent_dom_drivers}). 
Fire is an important Earth system process that is dependent on vegetation, climate, 
and anthropogenic activities. 
CESM2 incorporated a process-based fire model, which contains three 
components, namely, fire occurrence, fire spread, and fire impact \citep{Lewis_Firemodel_2013}. 
The interannual variability of agricultural fires is largely dependent 
on fuel load and harvesting; deforestation fires included fires due to natural 
and anthropogenic ignitions, caused by deforestation, land-use change, and dry climate. 
Peat fires usually occur in the late dry season and are strongly controlled by climate. 
The current version of the fire model reasonably simulates burned area, fire seasonality, 
fire interannual variability, and fire emissions \citep{Lewis_Firemodel_2013}. 
As fires are controlled by soil moisture, temperature, and wind, the attribution of NBP 
extremes to fires could also include the NBP extremes that are driven by inadequate 
soil moisture and hot temperatures.
Therefore, the total number of fire events could be larger, and recovery after 
such fire driven extremes could be much longer.  

Hot temperatures over long periods
tend to reduce ecosystem production and enhanced terrestrial respiration,
causing a large reduction in NBP \citep{Pan_2020_CarbonClimateExtremes}. 
Leaf photosynthesis depends on the RuBisCO-limited rate of carboxylation,
which is inversely proportional to the $Q_{10}$ function of 
temperature in the model \citep{Lawrence_2019_CLM5}. 
\citet{Hubau_2020_asynchronous} found that with increasing temperatures 
and droughts, tree growth was reduced and could offset earlier productivity gains. 
Conversely, warm temperatures in the northern high latitudes cause an 
increase in carbon uptake due to reduced snow cover and optimal temperature for photosynthesis. 
Increased warming at northern high latitudes could lead to hot temperature-related
 hazards and alter temperature--carbon interactions, which is discussed in Section~\ref{sec:tas_hig_lats}. 


Rising CO\textsubscript{2} emissions drive high temperature in 
the tropics and have the potential to hinder photosynthesis and 
vegetation growth, further discussed in Section~\ref{sec:extremes_in_tropics}. 
The changes in atmospheric circulation patterns might also influence 
the precipitation cycle, resulting in longer dry spells and 
increased fire risks with stomatal closure 
\citep{Frank_ClimateExtremes_CC_2015, Langenbrunner_Amazon_prcp_2019}. 
The second-largest negative NBP extremes were 
experienced by arid and semi-arid regions with mostly grasslands 
(Figure~\ref{f:Regions_NBP_TCE_extremes}(a)). 
Several studies conclude that soil moisture 
causes an increase in dry days and have a significant
negative effect on the carbon cycle driven by 
increasing droughts in arid, semi-arid, and dry temperate 
regions \citep{Frank_ClimateExtremes_CC_2015, 
Zscheischler_2014_GPP_extremes, Pan_2020_CarbonClimateExtremes}. 
The regions of South Africa, Central America, and 
Northern Australia witness the largest NBP extremes driven by fire. 
Extremes in the Amazon region were dominated by fire, soil moisture, 
and temperature in the 21\textsuperscript{st} century. 

\subsection{Compound Effect of Climate Drivers}
\label{sec:results_compound_effect} 

\begin{figure}
 \centering
 \includegraphics[trim = {0.1cm 0.1cm 0.1cm .15cm},clip,width=0.95\columnwidth]{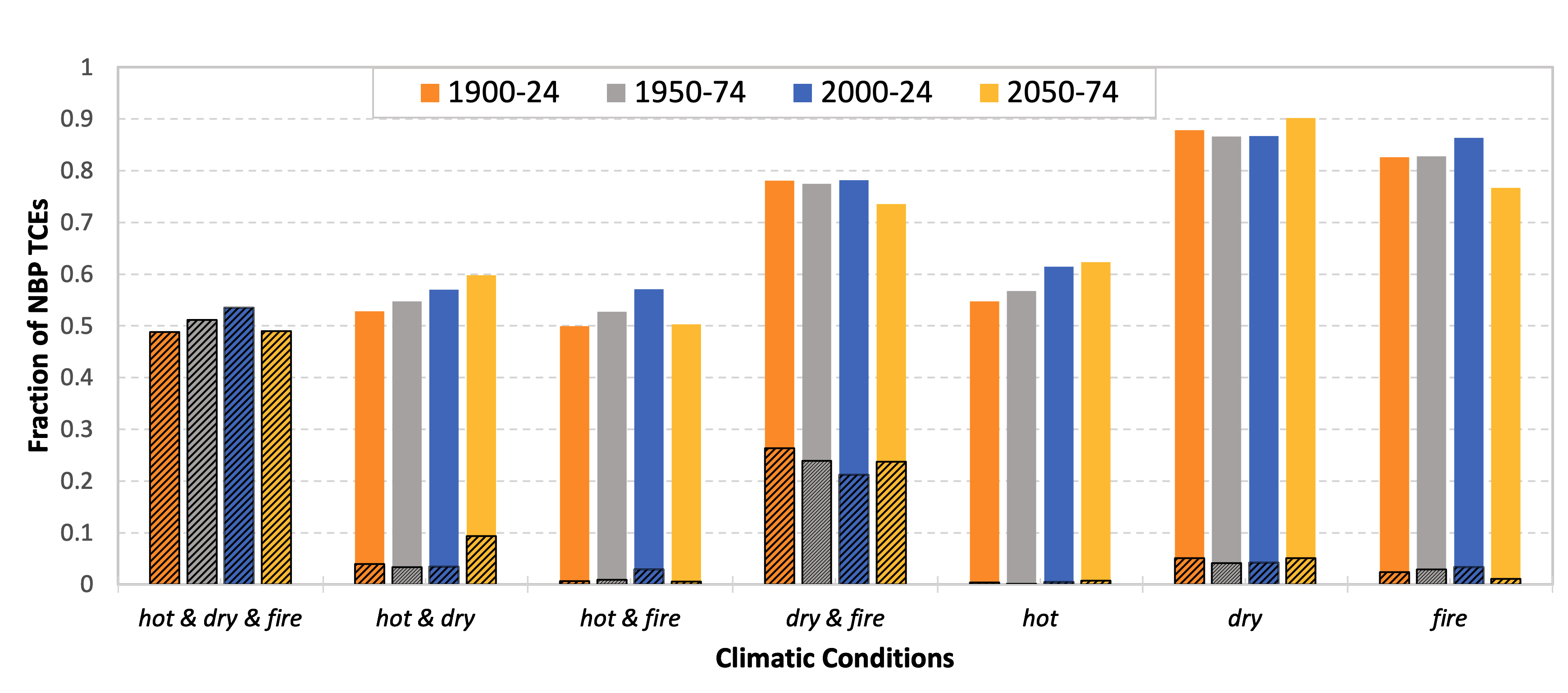}
 \caption{Fractional distribution of carbon cycle time-continuous extremes (TCEs) driven by compound climate drivers at lag of 1 month. 
 The unhatched and hatched bars represent the mutually inclusive and exclusive compound and individual climate drivers, respectively.
 The exclusive climate drivers are always less than or equal to mutually inclusive drivers. 
 The different colored bar represents following periods: 1900--24, 1950--74, 2000--24, and 2050--74 (\textit{from left to right bar}). 
 Most carbon cycle extremes are driven by interactive effect of climate drivers.}
\label{f:cesm2_nbp_compound_drivers}
\end{figure}

The interactive effect of multiple climate drivers could lead to 
devastating ecological consequences as compound extremes often 
have a larger impact on the carbon cycle than the aggregate response 
of individual climate drivers \citep{Zscheischler_Compound_2018, 
Ribeiro_2020_Crop, Pan_2020_CarbonClimateExtremes}. 
We used three broad classes of climate drivers, namely, moisture 
(dry vs. wet), temperature (hot vs. cold), and fire to study their compound effect. 
At most grid cells, NBP extremes were either positively correlated 
with anomalous precipitation and/or soil moisture, and/or 
negatively correlated with temperature and/or fire. 
Figure~\ref{f:cesm2_nbp_compound_drivers} shows the compound climate drivers,
both mutually exclusive and inclusive, that control NBP extremes over time.
Mutually inclusive climate drivers represent the simultaneous occurrence of 
various climatic conditions that drive extreme events in NBP.
Mutually exclusive climate drivers are those climatic conditions 
that do not occur at the same time to cause an extreme event.
For example, if an extreme event in NBP is driven by both
\textit{hot} and \textit{dry} conditions, 
the mutually exclusive climate driver is only \textit{hot \& dry} 
and the mutually inclusive drivers are \textit{hot}, \textit{dry}, and \textit{hot \& dry}.

The largest fraction, about 50\%, of total NBP TCEs were attributed to 
the combined effect of \textit{hot $\&$ dry $\&$ fire} events 
(Figure~\ref{f:cesm2_nbp_compound_drivers}). 
This implies that every other large extreme event associated with anomalous loss 
in biospheric productivity was driven by the interactive effect of water limitation, 
hot days (heat waves), both of which together could trigger fire and rapid loss of carbon. 
The second strongest exclusive compound driver was \textit{dry $\&$ fire}, causing about 25\% of extremes. 
With increasing climate warming, the number of NBP extremes driven by
hot and dry climatic conditions have increased, with about 10\%
extremes driven exclusively by \textit{hot $\&$ dry} events during
2050--74. 

Although the negative impact of water limitation (\textit{dry}) on NBP extremes was 
the highest (driving inclusively about 90\% of all NBP extremes), 
rising atmospheric CO\textsubscript{2} concentration and climate change led to an increasing number, 
54\% during 1900--24 to 62\% during 2050--74, of NBP extremes driven inclusively by 
\textit{hot} climatic conditions (Figure~\ref{f:cesm2_nbp_compound_drivers}).
For the same periods, extremes driven inclusively by \textit{hot $\&$ dry} rose by 8\%.



The effect of rising temperature on vegetation growth and carbon uptake is 
dependent on the geographical location. 
\citet{Pan_2020_CarbonClimateExtremes} found that net ecosystem production 
had a negative sensitivity to warming across 81\% of the global 
vegetated land area during 2007--18 and only the higher latitudes 
and Tibetan Plateau (TIB) had a positive sensitivity of NBP to temperature.
Similarly, \citet{Marcolla_2020_NBP} found a positive sensitivity of NBP to air 
temperature in higher latitudes and negative sensitivity in the tropics.
Since the tropics have the largest standing biomass and high latitude regions 
have the largest stored carbon, understanding the contrasting impacts of climate change across 
these regions is important to understanding climate--carbon feedbacks.
The next two sections will briefly discuss the changing characteristics 
of extremes in the tropics and at high latitudes.

\subsection{Increasing temperature sensitivity and weakening terrestrial carbon sink across the tropics}
\label{sec:extremes_in_tropics}

Observation-based studies have reported a decline in the rate of carbon
uptake in Amazonian forests, and similar declines in the African
tropics are expected in the future \citep{Hubau_2020_asynchronous}.
Over long timescales, the rising atmospheric CO$_{2}$ concentration
may not necessarily lead to an increase in plant biomass
\citep{walker2019decadal} as respiration losses outpace carbon uptake.
Increasingly frequent and stronger heatwaves, droughts, and fires due to climate
change are likely to cause 
the growth rate of NBP to flatten by the late 21\textsuperscript{st} century 
(Figure~\ref{f:ts_carbonfluxes_tropics}). 
They may lead to an eventual reduction in total stored carbon and 
a potential reversal for tropical vegetation from a net carbon sink to a carbon source.
Toward the end of the 21\textsuperscript{st} century (2075--99), most of the SREX regions (23
of 26) were dominated by negative NBP extremes
(Figure~\ref{f:Regions_NBP_TCE_extremes}(b)), especially in the tropical
regions (CAM, AMZ, NEB, WAF,
EAF, SAF, SAS, SEA, and NAU) (Figure~\ref{f:cesm2_nbp_net_uptake_reg}). 
During 2075--99, almost all tropical SREX regions (with the exception
of NAU) were dominated by negative NBP extremes. 

\begin{figure}
 \centering
 \includegraphics[trim = {0.1cm 0.1cm 0.1cm 0.1cm},clip,width=0.95\columnwidth]{
     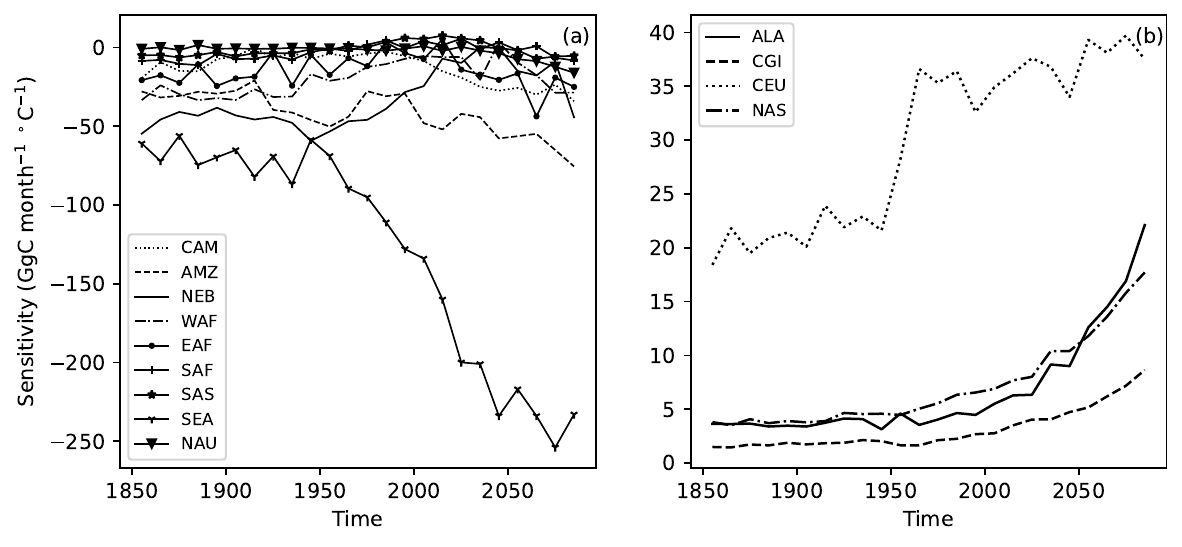}
 \caption{Changing temperature sensitivity of detrended anomalies in NBP 
 to detrended anomalies in surface temperature for 10 year time periods 
 at multiple SREX regions. The regions at low latitudes (a) have negative NBP 
 sensitivity to temperature anomalies and high latitudes, and (b) have positive 
 sensitivities.}
\label{f:sensitivity_tas_regional_nbp_10yrs}
\end{figure}

Rising temperatures and atmospheric CO$_{2}$ lead to an increasing trend
for gross primary production (GPP) and net primary production (NPP) across most of the tropics. However, they are often
also accompanied by increases in disturbance (such as droughts and
fire), inducing plant mortality and increases in heterotrophic respiration
that contribute to significant quantities of negative carbon fluxes from
the ecosystem. With the exception of AMZ and SEA that continued to witness
an increase in NBP in the model simulation, most of the tropical regions
showed a saturation or decline in NBP toward the end of
21\textsuperscript{st} century (Figure~\ref{f:ts_carbonfluxes_tropics}). 
Analysis of temperature trends across tropical regions showed a
significant trend toward warmer temperatures during warm (increase in
90\textsuperscript{th} quantile) as well as cool (increase in
10\textsuperscript{th} quantile) months of the year (Figure~\ref{f:tas_quant_tropics}).
Rising daily temperatures hinder net carbon uptake
by enhancing stomatal closure, under conditions of water stress, and 
increasing heterotrophic respiration (Figure~\ref{f:ts_carbonfluxes_tropics}).
The strength of 10 years of negative temperature sensitivity of NBP (see
Section~\ref{sec:sup_sensitivity}) increased over 
time (Figure~\ref{f:sensitivity_tas_regional_nbp_10yrs}(a)), 
suggesting an accelerated reduction in NBP growth with rising temperatures. 
The negative sensitivity values gradually increased from $-$20~GgC/month$\cdot^{\circ}$C 
to $-$33~GgC/month$\cdot^{\circ}$C for CAM, and 
$-$30~GgC/month$\cdot^{\circ}$C to $-$70~GgC/month$\cdot^{\circ}$C for AMZ during 1850--2100.
South-East Asia (SEA) saw the highest negative NBP sensitivity of $-$207~GgC/month$\cdot^{\circ}$C to temperature by the end of the 21\textsuperscript{st} century.
The possible reasons for the large difference in the NBP sensitivity for the region of SEA
compared to other tropical regions, e.g. AMZ, are the higher 
rate of decline in GPP sensitivity to 
temperature and the highest heterotrophic respiration (RH) sensitivity to
temperature of about 90~GgC/month$\cdot^{\circ}$C for the region of SEA. 
Our findings were consistent with Pan et al. (2020), who
analyzed seven Terrestrial Biosphere Models and 
found that the region of SEA had the largest negative NPP
sensitivity and positive RH sensitivity to temperature. 
Similar patterns were seen in other tropical regions, suggesting an
increasing negative temperature sensitivity of terrestrial ecosystem productivity to
carbon uptake in a warming world (Figure~\ref{f:sensitivity_tas_regional_nbp_10yrs}(a)).

\subsection{High latitude ecosystems can potentially become sources of carbon under warming climate}
\label{sec:tas_hig_lats}

High latitude ecosystems store large 
amounts of carbon below ground, and increasing exposure to warming and
disturbance pose the risk of release of stored soil carbon \citep{Marcolla_2020_NBP}
into the atmosphere.
Warmer temperatures at high latitudes create favorable conditions 
for longer growing seasons, enhanced plant growth and overall increases in greening. 
All high latitude regions (ALA, CGI, CEU, NAS) showed a trend of positive
and increasing NBP sensitivity to changes in air temperature
(Section~\ref{sec:sup_sensitivity}) over 1850--2100 (Figure~\ref{f:sensitivity_tas_regional_nbp_10yrs}).


While the overall impact of warming at high latitudes is expected to increase 
plant productivity and carbon uptake, high temperature anomalies increasingly 
induce negative NBP TCEs toward the end of the 21$^\mathrm{st}$ century.
The negative responses of NBP to warm air temperature anomalies were
found to occur most frequently during the summer months of July and
August.
For example, the 90\textsuperscript{th} quantile temperature increased from 
$13^{\circ}$C by $8^{\circ}$C to $21^{\circ}$C 
in the NAS region, while similar increases were observed for ALA, CGI and
CEU (Figure~\ref{f:tas_quant_highlat}(d)). 
With warming temperature trends, these periods of carbon losses
in response to temperature extremes have an oversized impact on the overall
carbon budget of high latitude ecosystems. Toward the end of the 21$^\mathrm{st}$
century, CGI and NAS showed strong declines in NBP, becoming a net source of
carbon. 



The accelerated warming of winter temperatures have large consequences 
for respiration losses in the Arctic and Boreal regions \citep{Natali_2019_ArcticResp, Jones_1998_ArcticResp, Commane_2017_ArcticResp}.
\citet{Natali_2019_ArcticResp} found that the total carbon loss from wintertime 
respiration in the Arctic was 60\% larger than the summer carbon uptake 
during 2003--17, driven primarily by higher soil and air temperatures.
Contrary to in situ observations, which show significant CO$_{2}$ emissions at subzero
temperatures, the current generation of process-based models shut off the 
respiration at subzero temperatures, thus underestimating the carbon losses 
during winter \citep{Natali_2019_ArcticResp}.
The simulation we analyzed showed a 1.7 times higher increase in winter air temperature
(10\textsuperscript{th} quantile) compared to summer air temperature
(90\textsuperscript{th} quantile) at high latitudes (Figure~\ref{f:tas_quant_highlat}).
For example, in NAS, the 10\textsuperscript{th} quantile 
temperature increased from $-25^{\circ}$C 
during 1900--24 by $14^{\circ}$C to $-11^{\circ}$C during 2075--99 (Figure~\ref{f:tas_quant_highlat}(d)). 
This enhanced rate of warming, especially during winter, resulted 
in rising wintertime total ecosystem (autotrophic and heterotrophic) respiration, 
turning some regions to net sources of carbon (Figure~\ref{f:ts_carbonfluxes_highlat}). 

Increases in warm and cold season temperatures induce a
potential risk of losing a carbon sink and an accelerated release 
of stored carbon into the atmosphere.
The increase in heterotrophic respiration is likely due to 
increased thaw of permafrost \citep{Turetsky_2020_permafrost}, 
a larger litter pool due to accelerated 
NPP, and higher microbial decomposition during the extended warm season.
As a result, the peak of NBP and NEP started to sharply decline toward the end of 21$^\mathrm{st}$ century.
The CGI region is expected to become a carbon source by the year 2100, and the 
NBP of the CEU region was gradually decreasing after 1975.
The NAS region has shown a reduction in total NBP during 
2075--2099, breaking the consistently increasing trend since 1850.
With accelerated rising winter temperatures (Figure~\ref{f:tas_quant_highlat}),
declining NEP and NBP (Figure~\ref{f:ts_carbonfluxes_highlat})
and underestimation of respiration in the current process models, 
the losses in carbon uptake in the Arctic
and at high latitudes in general are expected to be higher in the future.


\section{Conclusions}
\label{sec:conclusions}

The increasing frequency of climate change-driven extremes---such as fire,
drought, and heatwaves---have the potential to cause large losses of carbon
from terrestrial biomass and soils.
The increasing frequency and magnitude of negative NBP extremes
and saturation of NBP toward the end of the 21\textsuperscript{st}
century suggests that terrestrial ecosystems may increasingly lose the ability
to sequester anthropogenic carbon and ameliorate the impact of climate
extremes and change.
Under a changing climate, parts of the globe are expected to experience enhanced
vegetation growth and positive extremes in NBP; however, they are
far outpaced by the frequency and intensity of negative extremes and
associated losses in NBP. 
At the global scale, reductions in deforestation and enhanced CO\textsubscript{2}
fertilization lead to an increase in NBP.
The globally integrated NBP in the CESM2 reached a peak around 2070, 
followed by a large decline toward the end of the 21\textsuperscript{st}
century.
These losses in NBP were particularly large in
the carbon-rich tropical region, followed by arid and semi-arid regions of the
world.
During 2075--99, 23 out of 26 SREX regions were dominated by negative NBP 
extreme events, especially in tropical regions. 
The increasing intensity and magnitude of negative extremes in NBP toward
the end of the 21\textsuperscript{st} century and beyond could lead to
widespread declines in vegetation, loss of terrestrial carbon storage,
and increasingly turn terrestrial ecosystems into a net source of carbon.

Extremes in the carbon cycle, driven by the extremes in environmental
conditions, impact vegetation health and productivity. We analyzed
anomalies in three primary environmental drivers (\emph{hot, dry and fire}) of NBP extremes.
Negative anomalies in soil moisture, causing widespread droughts and water
stress in vegetation, were identified as the most dominant driver of
negative NBP extremes, affecting almost half of the grid cells experiencing NBP
extremes. 
The interactive and compounded impact of simultaneous anomalies in multiple
drivers have especially large impacts on vegetation productivity,
beyond the individual impacts of the variables. 
Extreme temperature anomalies compounded with dry conditions impact 
vegetation productivity, more than the sum of individual temperature and
moisture anomalies. 
They also increase the risk and occurrence of fires.
The compound effect 
of all three climate drivers (\emph{hot, dry and fire}) cause the largest fraction of NBP TCEs. 

In the tropics, the growth rate of NBP was decreasing, while the magnitude of 
negative extremes in NBP and the negative
temperature sensitivity of NBP was strengthening over time. 
Large standing carbon stocks (fuel load) with hot and dry 
climate (fire weather conditions) increases the fire risk 
and potential loss of carbon stock during negative NBP extremes.

In the northern high latitudes, accelerated warming leads to permafrost
thaw and release of belowground carbon, increasing the likelihood of 
reversal of the ecosystem to a net source of carbon over time.

This study analyzed climate-driven NBP extremes 
using one Earth system model, CESM2, from 1850 to 2100. 
Using only CESM2 simulations helped us to delve deeper into the 
climate-carbon feedbacks across different periods and spatial resolutions, 
as well as to identify model artifacts. 
However, the current study lacks comparison to observations, 
other Shared Socioeconomic Pathways, and other Earth System Models. 
Future work should use multi-model analysis to 
evaluate the agreement among different Earth system models about the 
magnitude, frequency, and spatial distribution of NBP extremes 
and their attribution to individual and compound climate drivers. 
Longer-term simulations are needed to analyze the climate--carbon 
feedback post-2100, when the difference between the rate of CO\textsubscript{2} 
emissions and terrestrial carbon uptake is expected to increase.


\codeavailability{Data analysis was performed in Python, and the
	analysis codes are publicly available on 
GitHub at \href{https://github.com/sharma-bharat/Codes_NBP_Extremes}{https://github.com/sharma-bharat/Codes\_NBP\_Extremes}.}

\dataavailability{
The data used here are from the CMIP6 simulations performed by the various modelling 
groups and available from the CMIP6 archive maintained by from Earth System Grid Federation (ESGF) 
\href{https://esgf-node.llnl.gov/search/cmip6}{https://esgf-node.llnl.gov/search/cmip6}.

At the CMIP6 archive site \href{https://esgf-node.llnl.gov/search/cmip6}{https://esgf-node.llnl.gov/search/cmip6} searching for a given model, 
a given experiment, and a given variable name will yield the link to the dataset that can be downloaded.
}




\appendix







\appendixfigures  

\appendixtables   

\clearpage
\section*{Supplementary}
\renewcommand{\thesection}{S\arabic{section}}
\renewcommand{\thetable}{S\arabic{table}}
\renewcommand{\thefigure}{S\arabic{figure}}
\setcounter{section}{0}
\setcounter{table}{0}
\setcounter{figure}{0}
\subsection{Calculation of temperature sensitivity of NBP} 
\label{sec:sup_sensitivity}
The sensitivity of NBP flux to surface air temperature is 
calculated using linear regression method \citep{Piao_2013_CC},
\begin{equation}
    \indent    NBP_{detrended} = b_0 + b_1 .~TAS_{detrended} + \epsilon
\end{equation}
where NBP$_{detrended}$ refers to the detrended monthly timeseries of NBP 
and TAS refers to the detrended monthly timeseries of surface air temperature.
The regression coefficient b$_1$ represent the apparent sensitivity of NBP to TAS;
b$_0$ is the fitted intercept; $\epsilon$ is the residue error in the 
linear regression.
The sensitivities of tropical and high latitudinal regions, as shown in 
Figure~\ref{f:sensitivity_tas_regional_nbp_10yrs}, has been calculated 
for consequentive 10 years periods starting from 1850 to 2100. 
The detrended timeseries of NBP and TAS for every SREX region were calculated 
by calculating the difference of area weighted mean and 10 year 
moving average of respective variables.
Positive temperature sensitivity to NBP signifies strengthening the impact of 
temperature on net biospheric carbon flux and vice versa.

\clearpage
\begin{table}
    \centering
    \caption{SREX Reference Regions}
    \begin{tabular}{c c}
    \hline
        Abreviation & Region's Full Name \\ [0.5ex]
        \hline
        ALA & Alaska/N.W. Canada  \\
        AMZ & Amazon  \\ 
        CAM & Central America/Mexico  \\ 
        CAS & Central Asia  \\ 
        CEU & Central Europe  \\ 
        CGI & Canada/Greenland/Iceland  \\ 
        CNA & Central North America  \\ 
        EAF & East Africa  \\ 
        EAS & East Asia  \\ 
        ENA & East North America  \\ 
        MED & South Europe/Mediterranean  \\ 
        NAS & North Asia  \\ 
        NAU & North Australia  \\ 
        NEB & North-East Brazil  \\ 
        NEU & North Europe  \\ 
        SAF & Southern Africa  \\ 
        SAH & Sahara  \\ 
        SAS & South Asia  \\ 
        SAU & South Australia/New Zealand  \\ 
        SEA & Southeast Asia  \\ 
        SSA & Southeastern South America  \\ 
        TIB & Tibetan Plateau  \\ 
        WAF & West Africa  \\ 
        WAS & West Asia  \\ 
        WNA & West North America  \\ 
        WSA & West Coast South America  \\ 
        \hline
    \end{tabular}
    \label{t:srex_reference}
\end{table} 
\clearpage
\begin{figure}
 \centering
 \includegraphics[trim = {0.1cm 0.1cm 0.1cm .85cm},clip,width=0.95\columnwidth]{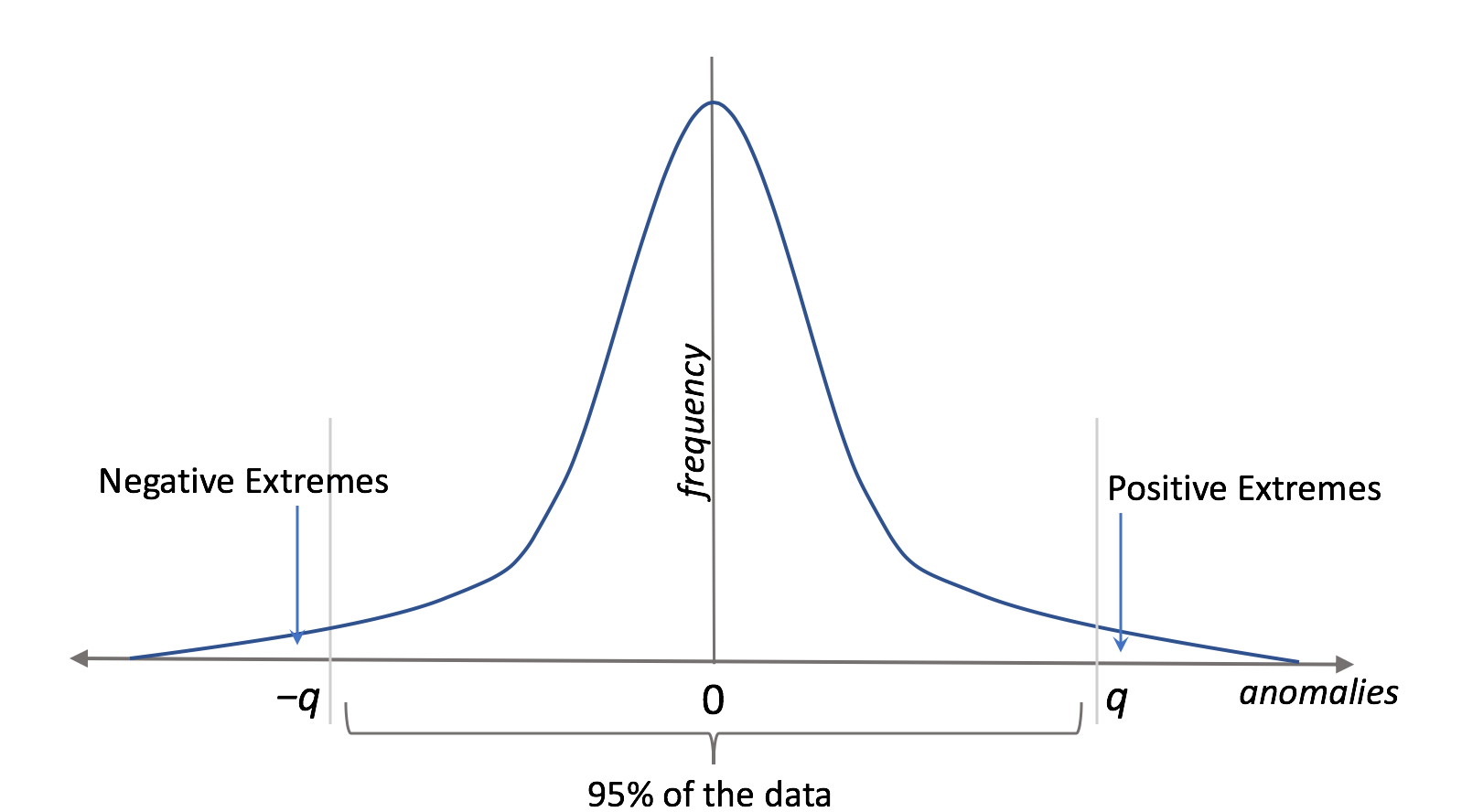}
 \caption{The schematic diagram representing the NBP extremes. 
 The threshold \textit{q} is set at 5$^{th}$
		 percentile in this study, such that 95\% of the NBP anomalies
		 lie within \textit{--q} and \textit{q}.}
\label{f:schematic_extremes}
\end{figure}
\clearpage
\begin{figure}
 \centering
 \includegraphics[trim = {0.1cm 0.1cm 0.1cm .25cm},clip,width=0.95\columnwidth]{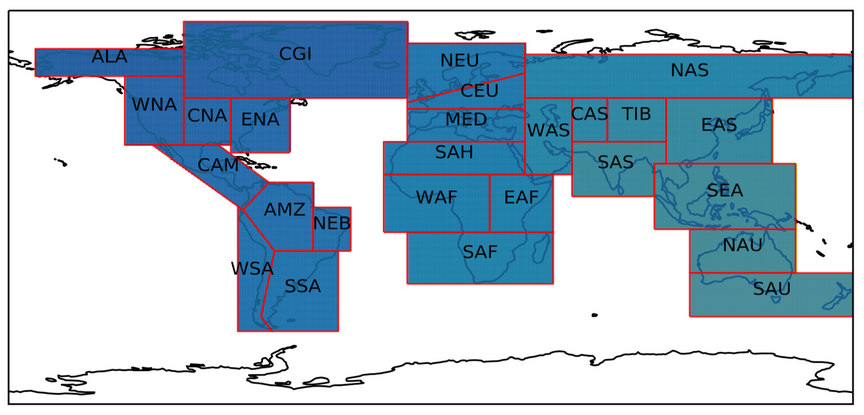}
 \caption{The spatial extent of SREX reference regions; abbreviations mentioned in the
		 Table~\ref{t:srex_reference}.}
\label{f:srex_regions}
\end{figure}
\clearpage
\begin{figure}
 \centering
 \includegraphics[trim = {0.1cm 0.1cm 0.1cm .95cm},clip,width=0.95\columnwidth]{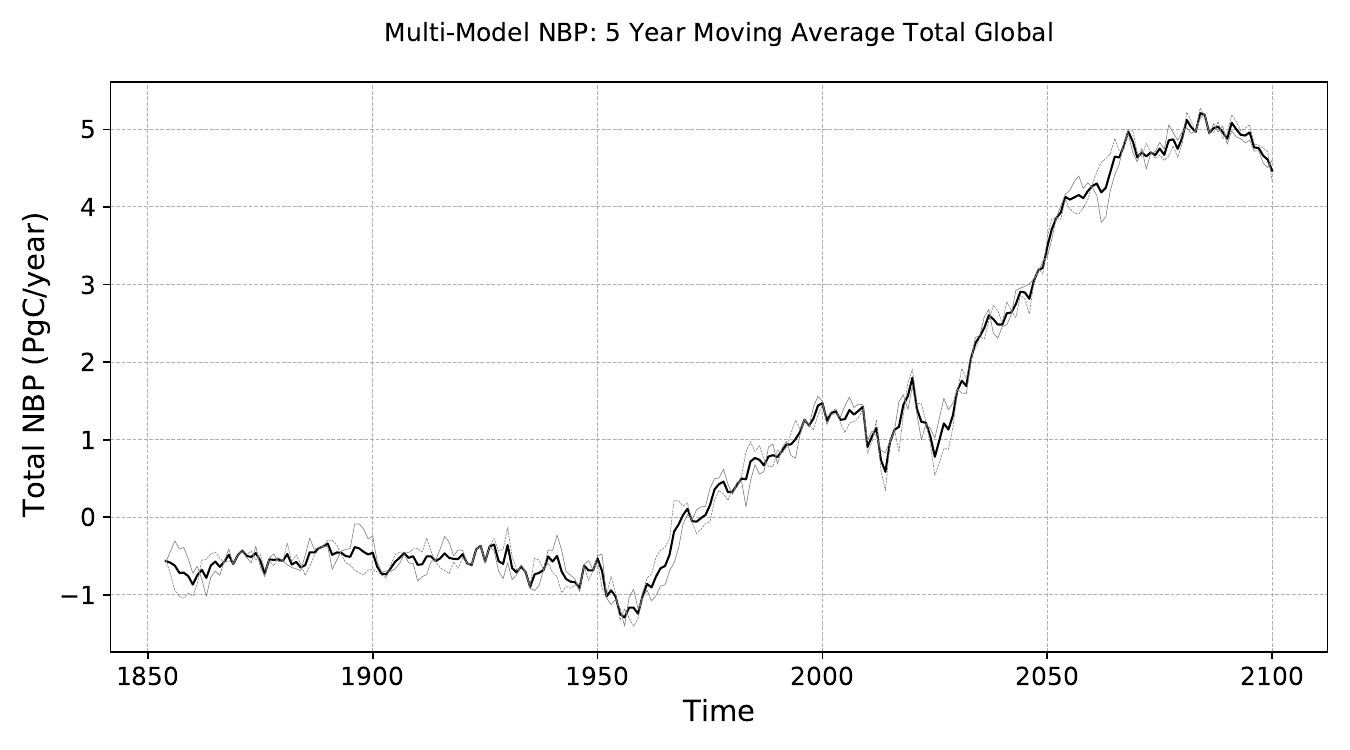}
 \caption{The timeseries of globally integrated 5~year rolling mean of NBP from 1850--2100 for 
 CESM2 ensemble members is shown in gray dashed lines. 
 The timeseries of globally integrated 5~year rolling mean of multi-ensemble mean is shown in black solid line.}
\label{f:nbp_cesm2_rm5yr}
\end{figure}
\clearpage
\begin{sidewaysfigure}
 \centering
 \includegraphics[trim = {.1cm .1cm 0.1cm 1.2cm},clip,width=0.95\columnwidth]{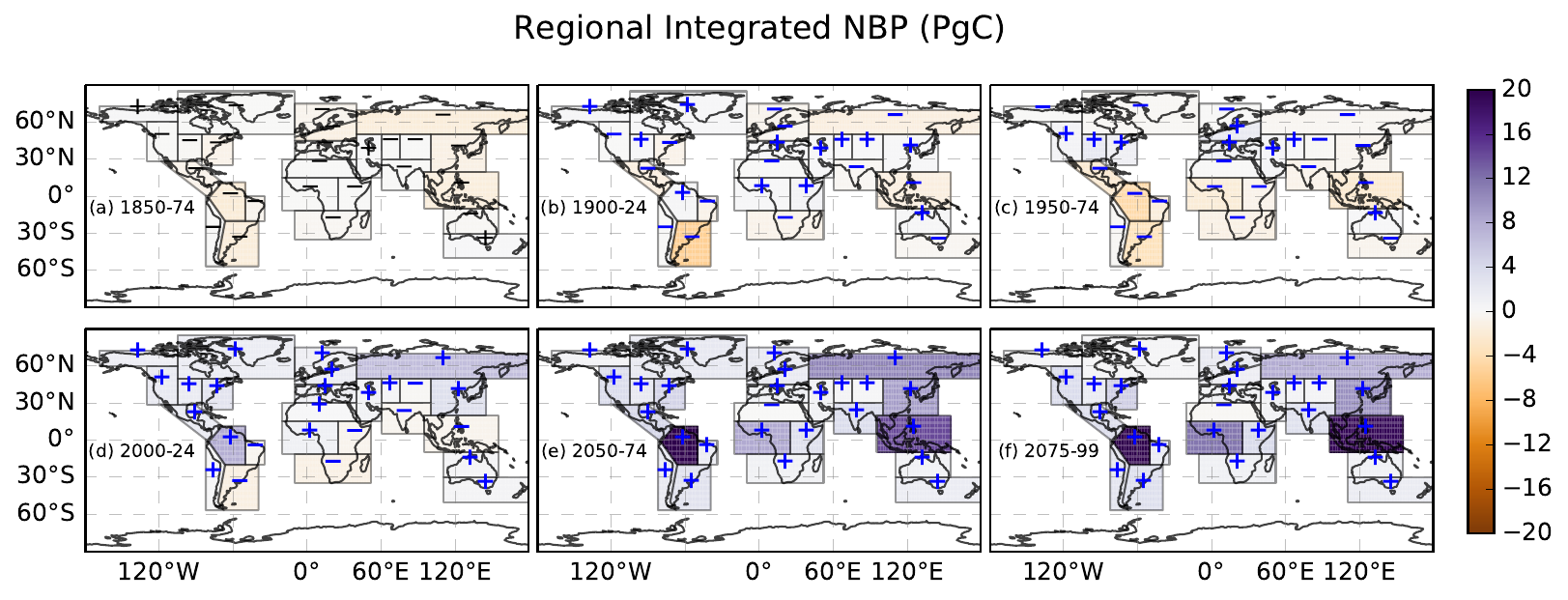}
 \caption{Total integrated NBP (PgC) for 25-year time windows for the period
	 1850--2100. Spatial distribution of integrated NBP (PgC) change over time: (a) 1850--74, (b) 1900--24, (c) 1950--74, (d) 2000-24, (e) 2050--74, and (f) 2075--99. Net increase in regional NBP or total carbon uptake is represented by purple color and `+' sign; net decrease is represented by orange color and `$-$' sign.}
\label{f:cesm2_nbp_reg_total}
\end{sidewaysfigure}
\clearpage
\begin{sidewaysfigure}
 \centering
 \includegraphics[trim = {0.1cm 6.1cm 0.1cm 7.25cm}, clip,width=0.95\columnwidth]{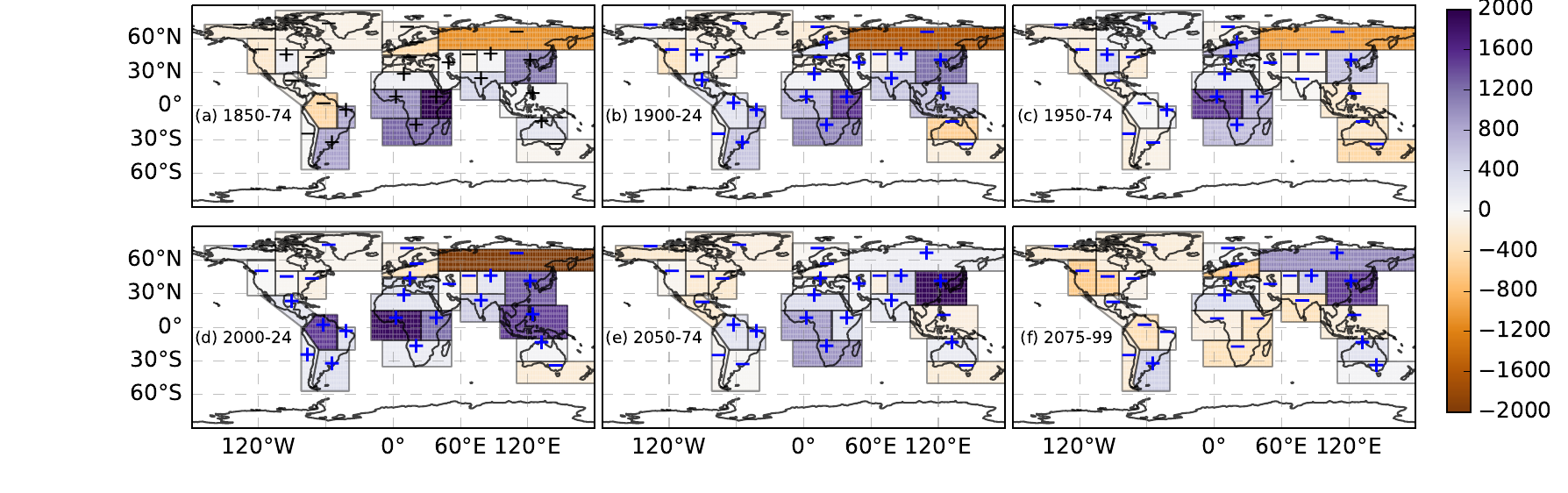}
 \caption{Frequency of positive~$ vs $~negative NBP extreme events across SREX regions. Purple color (`+'~sign) highlights the regions where frequency of positive NBP extremes events exceed negative NBP extremes; and brown color (`$-$'~sign) identifies regions where frequency of negative NBP extreme events exceed positive NBP extremes. Towards the end of 21st century, most tropical regions are dominated by frequent negative NBP extremes.}
\label{f:Spatial_NBP_fq_pos_neg_ext}
\end{sidewaysfigure}
\clearpage
\begin{figure}
 \centering
 \includegraphics[trim = {0.1cm 0.1cm 0.1cm .15cm},clip,width=0.65\columnwidth]{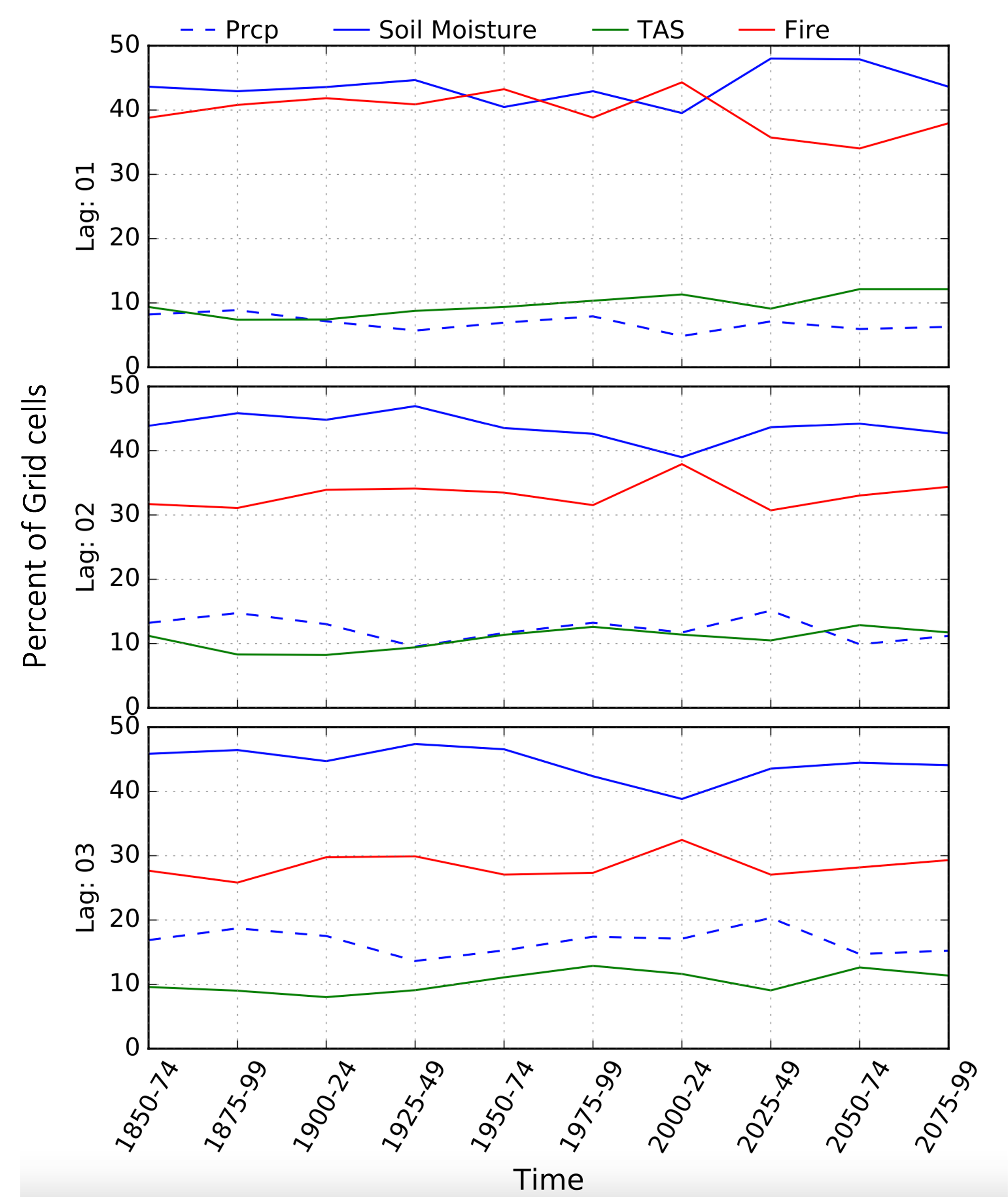}
 \caption{Percent distribution of number of grid cells with dominant climate drivers 
 causing time continuous carbon cycle extremes from 1850 to 2100 for every 25-year period. 
 The dominance of climate drivers is estimated by the absolute magnitude of correlation 
 coefficient (p~\textless~0.05) at lags of 1 (\textit{top}), 2 (\textit{middle}), and 3 (\textit{bottom}) months.}
\label{f:cesm2_nbp_percent_dom_drivers}
\end{figure}
\clearpage

\begin{figure}
 \centering
 \includegraphics[trim = {0.1cm 0.1cm 0.1cm .25cm},clip,width=0.95\columnwidth]{
     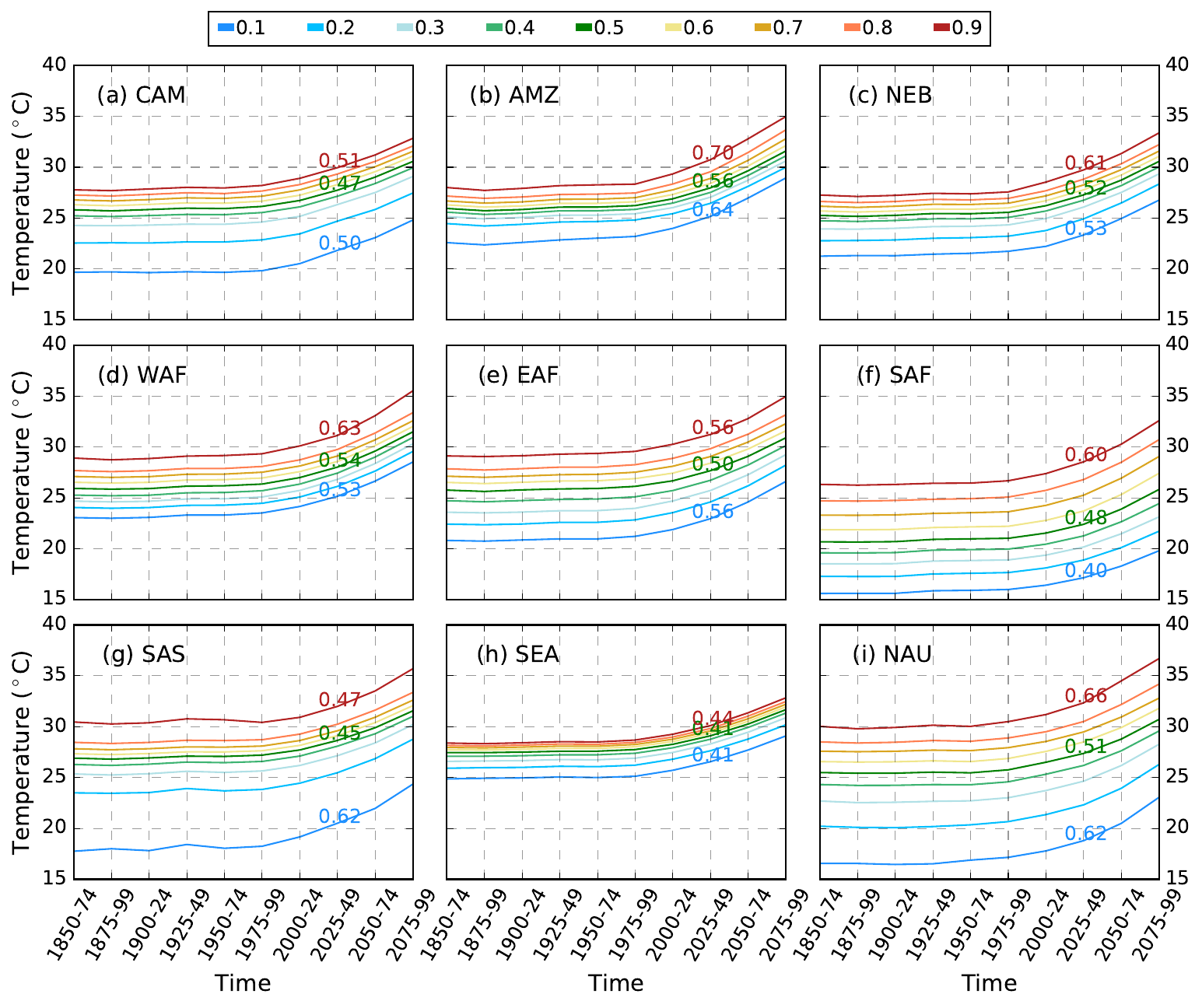}
 \caption{Change in area weighted average surface temperature (TAS) 
 at various quantiles in the 9 SREX regions in tropics for 25-year
	 windows from 1850--2100. The numbers shown in maroon, green, and blue 
	 in each subplot represent the rate of increase of temperature per decade 
	 ($^{\circ}$C/decade) for 90$^{th}$, median, and 10$^{th}$ quantile of 
	 temperatures, respectively.}
\label{f:tas_quant_tropics}
\end{figure}

\clearpage
\begin{figure}
    \centering
    \includegraphics[trim = {0.1cm 0.1cm 0.1cm .25cm},clip,width=0.95\columnwidth]{
        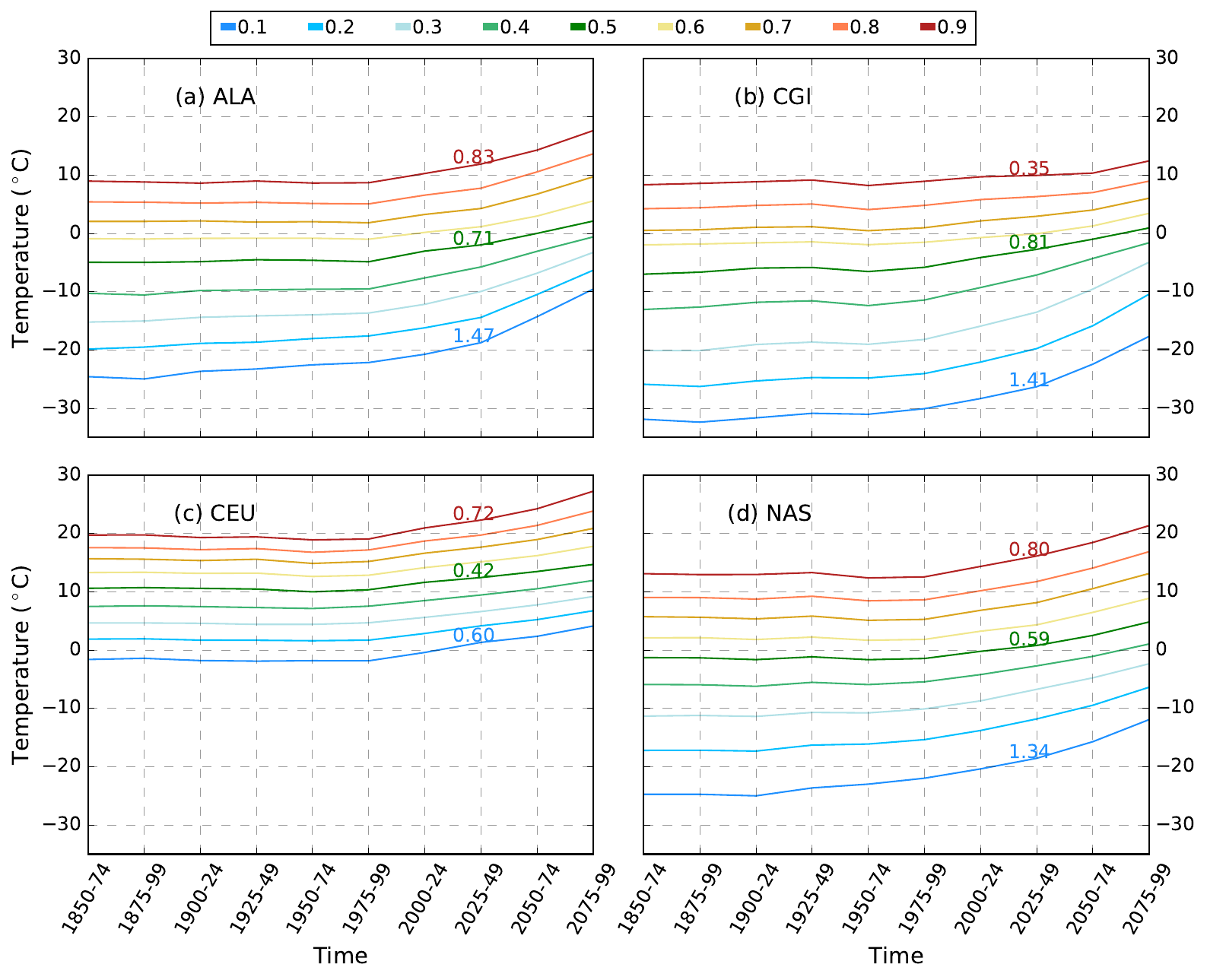}
    \caption{Change in area weighted average surface temperature (TAS) 
    at various quantiles in the 9 SREX regions at high latitudes for
		25-year windows from 1850--2100.
        The numbers shown in maroon, green, and blue 
        in each subplot represent the rate of increase of temperature per decade 
        ($^{\circ}$C/decade) for 90$^{th}$, median, and 10$^{th}$ quantile of 
        temperatures, respectively.}
   \label{f:tas_quant_highlat}
   \end{figure}
\clearpage
\begin{sidewaysfigure}
    \centering
    \includegraphics[trim = {0.1cm 0.1cm 0.1cm 0.25cm},clip,width=0.95\columnwidth]{
        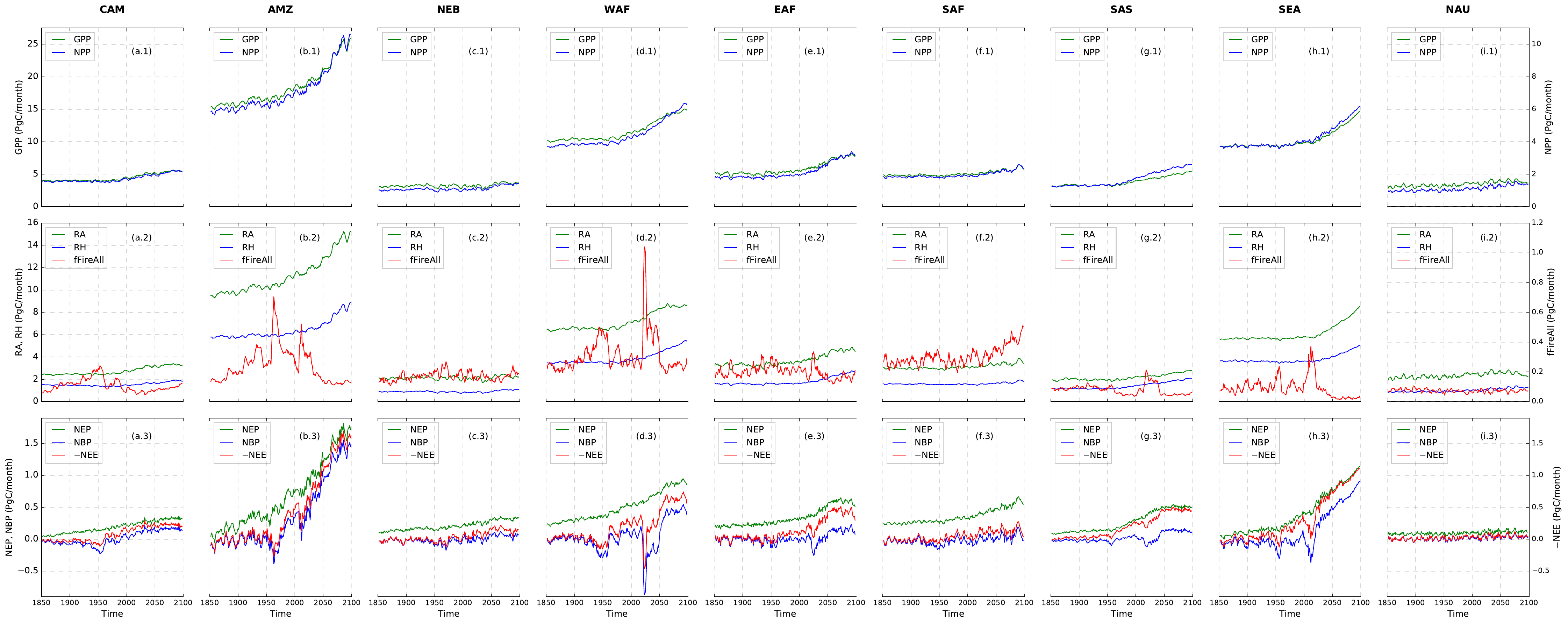}
    \caption{Timeseries of total carbon fluxes for the regions of (a) 
    Central America/Mexico (CAM), (b) Amazon(AMZ), (c) North-East Brazil (NEB), 
    (d) West Africa (WAF),(e) East Africa (EAF), (f) Southern Africa (SAF), 
    (g) South Asia (SAS), (h) Southeast Asia (SEA), and (i) North Australia (NAU) . 
    Row 1 for each region shows the 
    time series of total GPP (left y-axis) and NPP (right y-axis). 
    Row 2 shows RA and RH on left y-axis and fFireAll on right y-axis. 
    Row 3 shows NEP, NBP on left y-axis and $-$NEE on right y-axis.
    NEP is calculated by subtracting RH from NPP.
    NEP is surface net downward mass flux of carbon dioxide expressed 
    as carbon due to all land processes excluding anthropogenic land use change.
    $-$NEE has the consistent direction with the carbon flux such as NBP and NPP.}
   \label{f:ts_carbonfluxes_tropics}
   \end{sidewaysfigure}

\clearpage
\begin{sidewaysfigure}
    \centering
    \includegraphics[trim = {0.1cm 0.1cm 0.1cm 0.25cm},clip,width=0.95\columnwidth]{
        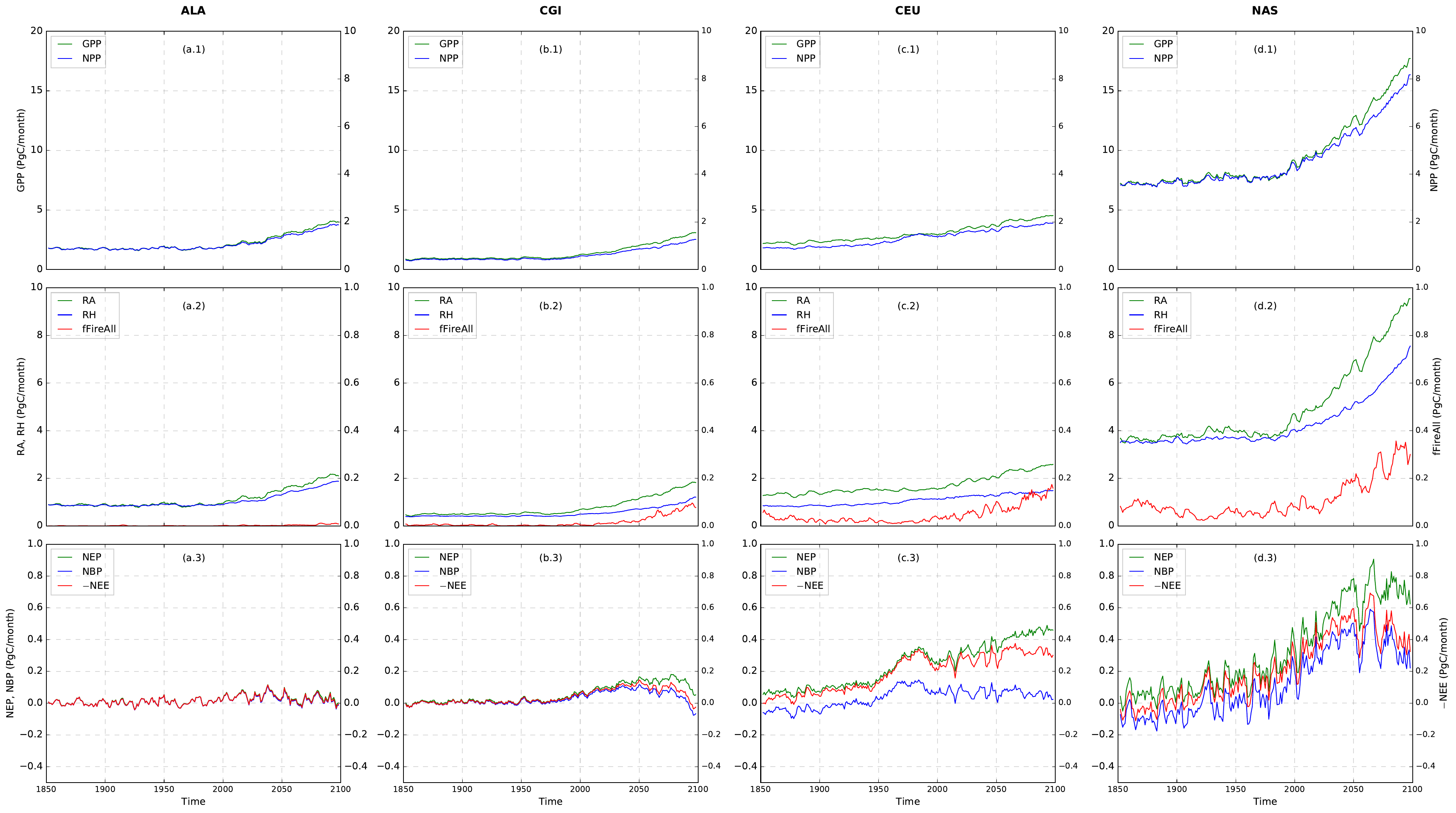}
    \caption{Timeseries of total carbon fluxes for the regions of (a) 
    Alaska, (b) Canada, Greenland, and Iceland, (c) Central Europe, 
    and (d) Northern Asia (NAS). 
    Row 1 of every region shows the 
    time series of total GPP (left y-axis) and NPP (right y-axis). 
    Row 2 shows RA and RH on left y-axis and fFireAll on right y-axis. 
    Row 3 shows NEP, NBP on left y-axis and $-$NEE on right y-axis.
    NEP is calculated by subtracting RH from NPP.
    NEP is surface net downward mass flux of carbon dioxide expressed 
    as carbon due to all land processes excluding anthropogenic land use change.
    $-$NEE has the consistent direction with the carbon flux such as NBP and NPP.}
   \label{f:ts_carbonfluxes_highlat}
   \end{sidewaysfigure}

\clearpage


\authorcontribution{BS designed the study with inputs from JK, FH, and AG. 
BS developed the statistical analysis methodology and codes, performed the data analysis, and wrote the 
manuscript with input from all co-authors.} 

\competinginterests{The authors declare that they have no conflict of interest.} 


\begin{acknowledgements}
    This research was supported by the Reducing Uncertainties in Biogeochemical Interactions through 
    Synthesis and Computation (RUBISCO) Science Focus Area, which is sponsored by the Regional and 
    Global Model Analysis (RGMA) activity of the Earth \& Environmental Systems Modeling (EESM) Program 
    in the Earth and Environmental Systems Sciences Division (EESSD) of the Office of Biological and 
    Environmental Research (BER) in the US Department of Energy Office of Science.
    This research used resources of the National Energy Research Scientific Computing Center (NERSC), 
    a U.S. Department of Energy Office of Science User Facility located at Lawrence Berkeley National 
    Laboratory, operated under Contract No. DE-AC02-05CH11231 for the Project m2467.

    We acknowledge the World Climate Research Programme, which, through its Working Group on Coupled Modelling,
    coordinated and promoted CMIP6. We thank the climate modeling groups for producing and making available 
    their model output, the Earth System Grid Federation (ESGF) for archiving the data and providing access, 
    and the multiple funding agencies who support CMIP6 and ESGF. 
    We thank DOE’s RGMA program area, the Data Management program, and NERSC for making this 
    coordinated CMIP6 analysis activity possible.
    
    The authors from ORNL are supported by the U.S. Department of Energy, Office of Science, Office of Biological and Environmental Research. ORNL is managed by UT-Battelle, LLC, for the DOE under contract DE-AC05-00OR22725.
\end{acknowledgements}






\bibliographystyle{copernicus}
\bibliography{bib_paper}

\end{document}